\documentclass[sigconf]{acmart} %

\usepackage{enumitem}
\usepackage[htt]{hyphenat} 
\usepackage{booktabs}
\usepackage{amsmath}
\usepackage{tabularx}
\usepackage{placeins}
\usepackage{microtype}
\usepackage{tikz}
\usetikzlibrary{arrows.meta,positioning,patterns}
\usepackage{pgfplots}
\usepackage{multirow}
\usepackage{graphicx}
\pgfplotsset{compat=1.18}
\DeclareMathAlphabet{\mathcal}{OMS}{cmsy}{m}{n}

\newcommand{\new}[1]{{\color{black}#1}}
\AtBeginDocument{%
  }

\copyrightyear{2026}
\acmYear{2026}
\setcopyright{cc}
\setcctype{by}
\acmConference[CCS '26]{Proceedings of the 2026 ACM SIGSAC Conference on Computer and Communications Security}{November 15--19, 2026}{The Hague, Netherlands}
\acmBooktitle{Proceedings of the 2026 ACM SIGSAC Conference on Computer and Communications Security (CCS '26), November 15--19, 2026, The Hague, Netherlands}
\acmDOI{10.1145/3830454.3846739}
\acmISBN{979-8-4007-2871-6/2026/11}

\begin{document}

\title{Grid Trouble in Paradise: Uncovering Vulnerable Distributed Energy Resources and Their Grid-Level Risks} %

\author{Anna Raymaker}
\affiliation{
  \institution{Georgia Institute of Technology}
  \city{Atlanta}
  \state{GA}
  \country{USA}}
\email{araymaker3@gatech.edu}

\author{Samuel Talkington}
\affiliation{
  \institution{Georgia Institute of Technology}
  \city{Atlanta}
  \state{GA}
  \country{USA}}
\email{talkington@gatech.edu}

\author{Zeezoo Ryu}
\affiliation{
  \institution{Georgia Institute of Technology}
  \city{Atlanta}
  \state{GA}
  \country{USA}}
\email{zeezooryu@gatech.edu}

\author{Richard Asiamah}
\affiliation{
  \institution{Georgia Institute of Technology}
  \city{Atlanta}
  \state{GA}
  \country{USA}}
\email{rasiamah3@gatech.edu}

\author{Emad Abukhousa}
\affiliation{
  \institution{Georgia Institute of Technology}
  \city{Atlanta}
  \state{GA}
  \country{USA}}
\email{emadak@gatech.edu}

\author{Betelihem Ashebo}
\affiliation{
  \institution{Georgia Institute of Technology}
  \city{Atlanta}
  \state{GA}
  \country{USA}}
\email{bashebo3@gatech.edu}

\author{Animesh Chhotaray}
\affiliation{
  \institution{Georgia Institute of Technology}
  \city{Atlanta}
  \state{GA}
  \country{USA}}
\email{achhotaray3@gatech.edu}

\author{Daniel K. Molzahn}
\affiliation{
  \institution{Georgia Institute of Technology}
  \city{Atlanta}
  \state{GA}
  \country{USA}}
\email{molzahn@gatech.edu}

\author{Frank Li}
\affiliation{
  \institution{Georgia Institute of Technology}
  \city{Atlanta}
  \state{GA}
  \country{USA}}
\email{frankli@gatech.edu}

\author{Saman Zonouz}
\affiliation{
  \institution{Georgia Institute of Technology}
  \city{Atlanta}
  \state{GA}
  \country{USA}}
\email{szonouz6@gatech.edu}

\author{Raheem Beyah}
\affiliation{
  \institution{Georgia Institute of Technology}
  \city{Atlanta}
  \state{GA}
  \country{USA}}
\email{ab207@gatech.edu}

\renewcommand{\shortauthors}{Raymaker et al.}

\begin{abstract}

Grid-connected solar distributed energy resources (DERs), such as solar inverters and monitoring platforms, have been deployed at unprecedented scale over the past few years, with global solar capacity more than doubling since 2022. To support monitoring and control, many of these systems are Internet-connected and configured by installers or end users, yet the real-world scale of their Internet exposure and the implications for power grid operation remain poorly understood.

In this paper, we present an Internet-scale evaluation of exposed and vulnerable solar DER infrastructure, and assess the risk that compromised DERs can pose to energy grids. We develop a method for accurately identifying solar DERs from Internet scanning data, and discover a diverse population of over 66,000 Internet-exposed solar DERs. We detect that at least 10,000 of these DERs may have known CVEs, such as unauthenticated monitoring and control endpoints. To assess the risk that these vulnerable DERs pose to a power grid, we use an electric grid network for Oahu, Hawaii, established and used by the power system research community, and conduct a power system analysis. Our evaluation shows that by compromising exposed DERs, attackers can cause voltage and line flow violations across multiple locations in the Oahu network, resulting in a range of consequences from degraded power quality to damaged power system components to power outages. Ultimately, our work brings to light the emerging threat of grid-connected DERs, and provides directions for improving energy security.

\end{abstract}



\maketitle

\section{Introduction}
\label{sec:intro}
 In December 2025~\cite{cert_pl_2025}, a coordinated cyberattack targeted Poland’s renewable energy sector, exploiting Internet-exposed control interfaces to disrupt distributed energy systems across the country. Rather than breaching centralized utility infrastructure, attackers exploited known vulnerabilities in edge-facing devices. This marked a shift from past attacks \cite{dragos2023electrum,cisa_ukraine2015,whitehead2017ukraine} on the Ukrainian power grid, which focused on high-value control centers, to a distributed attack model aligned with the grid’s evolution toward distributed energy resources (DERs)\footnote{DERs: small-scale grid-connected assets including solar inverters, batteries, and the meters and gateways used to monitor and control them.}~\cite{SolarDER-DOE}. Post-incident analysis of the attack in Poland estimates that a broader compromise could have removed over a gigawatt of generation capacity simultaneously~\cite{dragos2023electrum}. Such a sudden loss during peak demand could have triggered frequency instability and cascading failures, similar to those observed during the 2025 Iberian grid collapse \cite{IberianGridCollapse-wef}. The incident illustrates a recurring pattern in power-grid attacks: \textit{exploiting known vulnerabilities in reachable devices to induce system-wide disruption}.

The attack on DERs in Poland coincides with their rapid global expansion, particularly solar DERs. Between 2022 and 2024, global solar capacity doubled to 2 terawatts, driven by widespread adoption of grid-connected inverters and batteries. Unlike centralized generation, DERs inject power directly into distribution networks and are frequently managed via cloud dashboards, web interfaces, or vendor APIs. This architecture introduces new cyber–physical vulnerabilities, as devices (e.g., solar inverters) that directly affect grid behavior are now reachable from the public Internet~\cite{geller_exposed,freedman_exposed,dersec_2025}. The security implications are increasingly evident. In just two years, over 129 CVEs have been issued for solar DER platforms covering vulnerabilities in firmware, web dashboards, and control APIs~\cite{dersec_2025,nerc_whitepaper,hamilton_gridsecurity}. Yet despite the expanding threat landscape, two critical gaps remain: (a) \new{no reproducible, validated methodology exists for discovering the heterogeneous DER population at scale,} and (b) we lack a grounded understanding of the \textit{worst-case grid consequences} if such devices, especially solar inverters that directly control power generation, are compromised. In this paper, we address both challenges.

Prior work has made progress along two separate axes. Internet-wide scans have exposed vulnerabilities in cyber-physical systems~\cite{durumeric2015search,mirian2016internet,sasaki2022exposed,pickren2024release} but have not \new{systematically characterized solar DER exposure or connected it to grid impact}. Power-systems research has modeled the destabilizing potential of high-wattage load manipulation~\cite{soltan2018blackiot,shekari2022madiot,rodriguez2025madiot} but assumes hypothetical adversaries with control over millions of devices. Neither line addresses the central question for solar DERs: what can attackers actually do with the inverters exposed today? Answering this requires modeling power-supply-side attacks against the empirically exposed population -- fundamentally different from prior demand-side analyses, since compromised inverters control both active and reactive power and require \textit{orders of magnitude} fewer devices.

In practice, measuring DER exposure is challenging due to fragmented vendors, device roles, and installer-driven deployments that rely on non-standard, vendor-specific web interfaces. This heterogeneity produces noisy scanner artifacts that make it difficult to distinguish operational DER devices from generic web infrastructure at scale. Translating measured exposure into grid-level impact is similarly constrained by the limited availability of realistic distribution-network models with detailed DER placement.

These challenges and research gaps motivate the questions:\\
\noindent\textbf{RQ1:} How are solar DER platforms exposed on the public Internet, and what security-relevant control surfaces are reachable?\\
\noindent\textbf{RQ2.} What grid-level consequences could an attacker induce by manipulating these devices?

To answer these questions, we combine Internet-scale measurement with power-system analysis. We develop a cross-vendor discovery pipeline that identifies exposed solar DER platforms across Internet scan data. To filter non-device infrastructure at scale, we use an LLM to label a training set and a supervised classifier to apply those labels across 1.8M hosts. We then characterize reachable interfaces, map them to attacker capabilities, and simulate worst-case attacks on a realistic power grid model.

We identify 66,379 solar DER hosts spanning over 150 countries. Exposure is dominated by web-based management interfaces that enable device configuration and control, and at least 11,826 hosts are associated with known vulnerabilities. To our knowledge, this is the largest population of Internet-exposed cyber-physical devices identified by a measurement study, exceeding prior work on PLCs~\cite{pickren2024release} and EV charging systems~\cite{nasr2023chargeprint} \emph{by a factor of two or more}.

Grounding this exposure in a power-system context, we conduct the first measurement-driven grid impact analysis in the security community. Using exposed inverters in Oahu, we show that coordinated manipulation of just 41.1 MW of generation is sufficient to drive 18 of 37 buses more than 10\% below nominal voltage and overload \new{6} of 89 transmission lines on a realistic grid model, conditions that can damage motors and transformers and force protective disconnections that lead to outages. Unlike prior (power) demand-side attacks that require coordinating millions of high-wattage IoT devices~\cite{shekari2022madiot,soltan2018blackiot}, supply-side compromise gives attackers direct control over both active and reactive power. This enables voltage manipulation that demand-side attacks cannot achieve, with orders of magnitude fewer devices.

Our findings demonstrate that insecure DER interfaces are widespread and operationally consequential. They provide adversaries with an expanding and poorly secured attack surface that warrants immediate attention from vendors, operators, and regulators.

In summary, \textbf{our contributions} are as follows:\\
$\bullet$ \new{A} measurement-driven power-system analysis that maps empirically observed Internet-exposed solar DERs into attacker capabilities and quantifies resulting grid-level violations using a realistic power grid model.\\
$\bullet$ A global, Internet-scale measurement of 66,379 exposed solar DER hosts, produced by a scalable discovery pipeline that combines deterministic fingerprinting, LLM-assisted labeling, and supervised classification to separate operational DERs from Internet noise across heterogeneous vendor ecosystems.\\
$\bullet$ \href{https://gridtrouble.xyz}{https://gridtrouble.xyz}, a public tool that simulates and visualizes the maximum grid impact achievable on Oahu \emph{using actual, currently exposed DERs from our measurement}. Users can select inverter subsets, adjust control settings, and watch the impacts change.

\section{Threat Model, Related Work, and Overview}
In this section, we present our threat model, review prior work on Internet-wide discovery of networked devices and on the system-level impact of large-scale device compromise, and give an overview of our measurement and power-analysis approach.

\subsection{Threat Model}
We consider an Internet-based adversary targeting solar-DER infrastructure such as rooftop and small-commercial solar inverters, monitoring gateways, and fleet, or installer-managed platforms, through publicly reachable interfaces. Consistent with known attacks~\cite{cert_pl_2025} against DERs, we consider the adversary operates without physical proximity or on-premises access and interacts with devices solely via the public Internet, including web management interfaces, monitoring dashboards, installer portals, and exposed industrial control protocols (e.g., Modbus TCP).  When authentication is weak or absent, the attacker may read or modify inverter configuration parameters such as power setpoints, operational modes, or device state, and may trigger actions including disconnection, reboot, or firmware upload. We infer the capabilities of the attacker only from scanner-visible interfaces and documented control functionality.

\subsection{Related Work}
\paragraph{\textbf{Internet-Wide Discovery.}}
Prior work has shown that Internet-wide scanning and indexing platforms such as Censys and Shodan enable large-scale measurement of exposed networked systems~\cite{durumeric2015search}. Early studies revealed widespread exposure of industrial control systems (ICS), often lacking authentication or basic access controls~\cite{mirian2016internet}. Subsequent work identified exposed ICS remote management interfaces and extended Internet-wide measurement techniques to other cyber-physical domains such as robotics and automotive systems~\cite{sasaki2022exposed,wu2025revealing,chen2024understanding,ueda2022internet,salazar2024tale}.

More recent research has introduced automated discovery pipelines. PLCHound~\cite{pickren2024release} showed that iterative, cross-protocol inference can substantially improve recall for programmable logic controllers (PLCs), while ChargePrint~\cite{nasr2023chargeprint} applied Internet-scale measurement to electric vehicle charging systems. These approaches succeed because their targets share structural regularities that discovery can key on: PLCs expose a small set of standardized industrial protocols (e.g., Modbus, EtherNet/IP, CODESYS) and \new{consistent product identifiers (e.g., recognizable web-management banners)} that generalize across vendors, and EV chargers commonly implement OCPP~\cite{OCPP}. DER ecosystems share \new{none of these regularities}, so existing discovery techniques do not readily generalize to DERs\new{; applying PLCHound's method~\cite{pickren2024release} to our DER population confirms this, returning few devices.}

\new{These structural differences also limit brand- and product-anchored discovery, as in Forescout's Shodan-based snapshot~\cite{geller_exposed}: it misses devices without a recognizable identifier (53.7\% of our validated hosts carry no vendor fingerprint; Section~\ref{sec:hetero}) and relies on scanner-provided device labels shown to be unreliable~\cite{pickren2024release}. Our pipeline instead recovers this heterogeneous, largely unbranded population. More fundamentally, prior industry and academic efforts characterize exposure but stop there. We are the first to bridge exposure and impact: connecting a measured, validated DER population to a quantified, grid-level consequence analysis (Section~\ref{sec:power-analysis}) is the central contribution of this work.}

\paragraph{\textbf{Large-Scale Device Compromise.}}
A complementary line of work has examined the consequences of large-scale compromise of networked devices. Studies of IoT botnets, most notably Mirai~\cite{antonakakis2017understanding}, showed how insecure ecosystems enable rapid, widespread device takeover. In the energy domain, BlackIoT~\cite{soltan2018blackiot,singer2023shedding,huang2019not} and MaDIoT~\cite{shekari2021mamiot,shekari2022madiot,rodriguez2025madiot,maiti2023targeted} showed that coordinated control of high-wattage IoT devices can destabilize power systems even in the presence of standard grid protections. Other work has shown that attacks on individual solar inverters can induce voltage and frequency disturbances~\cite{barua2020hall}.

Our work differs from this line in three ways. First, BlackIoT and MaDIoT are \emph{demand-side} attacks that manipulate load through compromised IoT devices; we study \emph{supply-side} attacks that compromise generation directly through exposed inverters. This distinction is consequential: demand-side attacks can only manipulate active power, while compromised inverters independently control both active and reactive power, enabling voltage manipulation that load-based attacks cannot achieve. Second, prior work assumes an adversary already controls a hypothetical population of millions of devices; we instead measure the population that is actually Internet-exposed and vulnerable and ask what an attacker controlling \emph{that} population can do. Third, supply-side compromise achieves comparable grid impact with orders of magnitude fewer devices than demand-side attacks require.

\subsection{Approach Overview}
\label{sec:overview}

Our approach combines Internet-scale measurement with power-system analysis to map DER exposure and quantify its grid-level consequences.

\paragraph{\textbf{Measurement pipeline (Sections \ref{sec:method} and~\ref{sec:exposure_results}).}}
Our measurement is anchored in publicly available Internet scan data from Censys~\cite{censys}, which provides protocol- and service-level observations of IPv4 and IPv6 hosts. We execute our discovery pipeline over two single-day Censys snapshots taken approximately four months apart. Snapshot~1 applies queries for 34 DER vendors and yields 166K candidate hosts; Snapshot~2 expands coverage to 96 vendors and yields 1.8M candidates. For each candidate, we extract scanner-visible metadata (service banners, open ports, TLS certificate fields, and HTTP response data) and, where web services are exposed, retrieve and parse associated web content. Using Snapshot~1, we produce a labeled dataset via LLM-assisted classification and deterministic refinement (Section~\ref{sec:labeling}). These labels train an interpretable low-latency classifier that scales DER identification to the 1.8M hosts in Snapshot~2 (Section~\ref{sec:scaled_classification}), yielding 66,379 validated DER hosts.

\paragraph{\textbf{Power-system analysis (Section~\ref{sec:power-analysis}).}}
To translate measured exposure into grid-level impact, we augment a validated synthetic transmission model of Oahu~\cite{tamu_electric_grids_website} with the exposed inverter population identified above. We isolate all 571 exposed and vulnerable DERs in Oahu and manually verify each for attacker-relevant exposures (e.g., unauthenticated management interfaces, firmware upload endpoints, or externally reachable control protocols) and aggregate them into compromisable units totaling 41.1~MW of generation capacity. We formulate worst-case attacks as adversarial AC optimal power flow problems: compromised inverters maximize voltage and line-flow violations, subject to physical inverter limits and the automatic response of non-compromised generators. This produces an upper bound on grid impact for the empirically observed attack surface.

\section{Methodology}

\label{sec:method}

\begin{figure*}[]
    \centering
    \includegraphics[width=\linewidth]{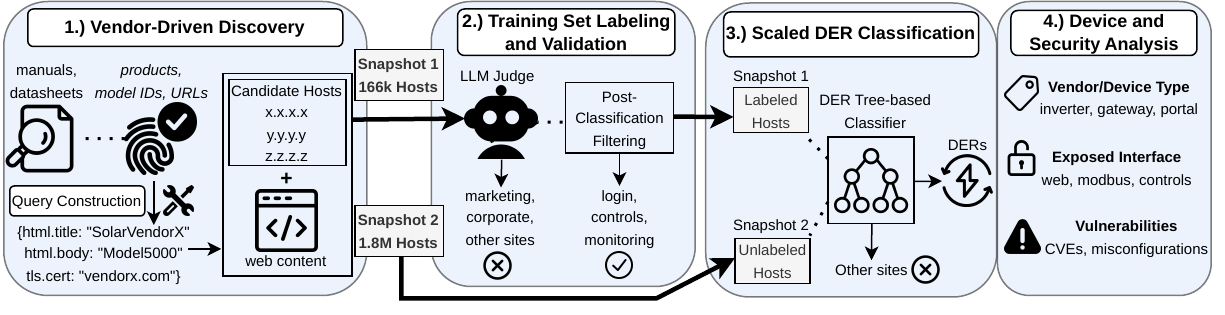}
    \caption{\textbf{Methodology overview for discovering and
    characterizing Internet-exposed DERs.}  We derive vendor-specific
    fingerprints from public documentation and construct deterministic
    Censys queries to identify candidate hosts. An LLM-based classifier
    labels a first snapshot (34~vendors), producing training data for a
    tree-based model that scales classification to a second, broader
    snapshot (96~vendors) taken four months later.  Validated DER hosts
    are then characterized by vendor, interface exposure, and security
    properties.}
    \label{fig:method_overview}
    \vspace{-1em}
\end{figure*}

We present a measurement pipeline for discovering and characterizing DER devices that are reachable from the public Internet. The pipeline proceeds in four stages (Figure~\ref{fig:method_overview}).  First, we derive vendor-specific fingerprints from public documentation and construct deterministic queries over Internet scan data to identify candidate hosts (\emph{Vendor-Driven Discovery}). Second, we apply an LLM-based classifier to a single-day Censys snapshot to produce labeled training data (\emph{Training-Set Labeling and Validation}). Third, we train a tree-based classifier on those labels and apply it to a second, larger snapshot taken approximately four months later to identify DER hosts at scale (\emph{Scaled DER Classification}). Finally, we characterize each validated host by vendor, interface exposure, and security properties (\emph{Device and Security Analysis}).

The two-snapshot design serves two purposes.  First, it enables
\emph{scalable classification}: the LLM produces high-quality  labels on Snapshot~1 (166K hosts) but is too slow for per-host inference on Snapshot~2 (1.8M hosts), so a lightweight supervised model trained on those labels handles the larger set.  Second, it enables a \emph{generalization test}: Snapshot~2 was collected approximately four months after Snapshot~1 and incorporates queries for 96 DER vendors compared to 34 in Snapshot~1, allowing us to evaluate whether the classifier can identify previously unseen brands and whether exposure patterns persist over time.

\subsection{Stage 1: Vendor-Driven Discovery}
\label{sec:discovery}

We systematically identify Internet-facing DER candidates by mining publicly available vendor documentation and translating the resulting fingerprints into deterministic Censys queries.

\noindent\textbf{Fingerprint Extraction.}
For each target vendor and monitoring platform
(Table~\ref{tab:brand_types} in the appendix), we scrape installation manuals, datasheets, and support documentation. From these sources, we extract product names and model numbers that appear in configuration or status interfaces (e.g., SMA \emph{Sunny Boy}, \emph{Sunny WebBox}), user-interface strings from documented dashboards and login pages, and portal domains used for remote monitoring or fleet management. Identifiers are retained only when they co-occur with solar or photovoltaic context in the source documentation; generic tokens without energy-specific qualifiers are discarded.  For vendors with common names (e.g., ABB, Delta, Siemens), identifiers are constrained to co-occur with DER-specific terms such as \texttt{solar inverter} or \texttt{photovoltaic} to prevent matches against unrelated industrial systems.  Minor textual variants are consolidated through exact matching
and edit-distance clustering.

Industry data indicates that the top ten global inverter vendors account for at least 86\% of worldwide PV inverter shipments~\cite{chopra_solar,sylvia_solar,fortune_solar}. Our initial vendor set of 34 covers the major manufacturers and commonly deployed monitoring ecosystems. For Snapshot~2, we expand coverage to 96~vendors by incorporating additional regional manufacturers, white-label products, and monitoring platforms identified during error analysis of Snapshot~1 results.

\noindent\textbf{Query Construction.}
Each fingerprint is translated into a Censys query that searches
scanner-visible fields where deployment identifiers are likely to
appear, including HTTP response titles and bodies, TLS certificate
subject fields, and service-level response data. Multi-field matching is necessary because deployed DER systems surface identifiers inconsistently across titles, page content, certificates, and auxiliary service metadata.  For vendors with distinctive identifiers (e.g., \emph{IQ~Gateway}, \emph{SUN2000}), fingerprints are matched directly as exact or near-exact strings. For ambiguous vendors, queries
conjoin vendor identifiers with energy-domain terms.

In addition to brand-specific queries, we construct a single
\emph{generic} discovery query from recurring DER-specific interface terms (e.g., \texttt{photovoltaic system}, \texttt{solar dashboard}) that appear across vendors once brand names are removed. This generic query targets white-label, rebranded, or otherwise unattributed devices. All queries are generated programmatically from the fingerprint database and stored prior to execution, ensuring reproducibility. Further details on queries and field-matching logic are in Appendix~\ref{app:query_details}.

\noindent\textbf{Snapshots and Coverage.}
We execute the discovery pipeline at two points in time. \emph{Snapshot~1} comprises 166K candidate hosts from a single-day Censys scan using 34~vendor-specific queries plus one generic query. \emph{Snapshot~2}, taken approximately four months later, incorporates 96~vendor queries and BigQuery-based full HTTP body search, yielding 1.8M candidate hosts. Vendor-specific queries identified nearly 50\% of validated DER hosts that would have been missed by generic queries alone, while producing 14.7\% false positives compared to 85.3\% for the generic query. \new{We also filter honeypots; see Honeypot Handling in Section~\ref{sec:scaled_classification} below for more details.}

\subsection{Stage 2: Training-Set Labeling and Validation}
\label{sec:labeling}

Snapshot~1 contains both operational DER devices and non-device artifacts such as vendor marketing pages, academic resources, and unrelated industrial infrastructure.  Accurately labeling 166K hosts requires a classifier that can generalize across diverse interface styles without a pre-existing labeled dataset.  We therefore apply an LLM in a zero-shot setting --- the only approach that can produce reliable labels at this scale without manual annotation of tens of thousands of hosts.

\noindent\textbf{LLM-Assisted Classification.}
For each host, an off-the-shelf large language model receives scanner-visible metadata (service banners, ports, protocols, TLS fields) and scraped web content (HTML titles, bodies, headers, and reachable paths) and classifies it as either a DER-related deployment or a non-DER artifact.  Simple keyword matching is insufficient for this task because distinguishing a page that \emph{is} a device from one that merely \emph{discusses} a device requires semantic understanding of page structure and content~\cite{NEURIPS2020_1457c0d6}.

\begin{table}[]
  \centering
  \caption{LLM-based device classification performance}
  \small
  \setlength{\tabcolsep}{5pt}
  \renewcommand{\arraystretch}{1}
  \begin{tabularx}{\linewidth}{@{}Xcccccc@{}}
    \toprule
    \textbf{LLM} & \textbf{CL} & \textbf{\#IPD} & \textbf{\#O} & \textbf{Precision} & \textbf{Recall} \\
    \midrule
    \textbf{Llama3.3:70b} & 8K & 67,293 & 98,710 & 58\% & 97\% \\
    Gemma3:27b & 8K & 86,222 & 79,781 & 34\% & 81\% \\
    Qwen3:32b & 8K & 32,631 & 133,372 & 60\% & 97\% \\
    Qwen3:32b & 128K & 31,808 & 134,195 & 58\% & 88\% \\
    \bottomrule
  \end{tabularx}

  \vspace{0.4em}
  \footnotesize
  CL = context length; \#IPD = identified potential DERs; \#O = other devices.\\
  Precision and recall are estimated from manually labeled samples of 50 predicted-DER and 50 predicted-non-DER hosts for each LLM (100 total).
  \label{tab:llm_comparison}
\end{table}

We evaluated four open-source LLMs spanning different architectures and context windows (8K--128K tokens), summarized in Table~\ref{tab:llm_comparison}. Each classifies hosts using a structured three-component prompt that separates invariant classification logic from per-host evidence (Figure~\ref{fig:prompt_structure}), detailed in Appendix~\ref{app:llm_refinement}. \new{We select \textbf{Llama~3.3:70b}, which identifies a larger set of candidate DER hosts than the Qwen-based models while exhibiting comparable true positive and negative rates. Retaining a larger candidate set here is desirable because subsequent stages conservatively eliminate residual false positives, so high recall matters more than precision at this step. We use open-weight models instead of commercial APIs, as they achieve sufficient screening performance while avoiding per-token inference costs over a corpus of this size (166K hosts in Snapshot~1).} Across all evaluated models, misclassifications are dominated by false positives rather than false negatives, reflecting the intended role of the LLM as a high-recall screening filter. Applying Llama~3.3:70b reduces the candidate set from 166K to 67K hosts.

\noindent\textbf{Deterministic Refinement.}
Manual inspection of LLM errors reveals that false positives cluster around a small number of recurring patterns: consumer-grade solar-powered cameras (e.g., Reolink), network management appliances, and academic pages that discuss power systems or expose industrial protocols without operational control surfaces. We encode these patterns as explicit suppression rules operating on scanner-visible characteristics (HTML titles, server headers, protocol combinations, vendor strings). This hybrid approach, LLM screening plus deterministic suppression, achieves 98\% precision on a manually validated random sample (Table~\ref{tab:classification_validation}). Full details of the error taxonomy, suppression rule construction, and final validation are provided in Appendix~\ref{app:llm_refinement}.

The output of this stage is a set of labeled DER (41,755) and non-DER (124,208) hosts from Snapshot~1, which serves as the training set for Stage~3.

\subsection{Stage 3: Scaled DER Classification}
\label{sec:scaled_classification}

Snapshot~2 contains 1.8M candidate hosts. Per-host LLM inference at this scale is impractical: classifying Snapshot~1 (166K~hosts) with Llama~3.3:70b alone required 634~GPU-hours across 17~jobs on H100/H200 GPUs, and evaluating all four candidate LLMs consumed 1,426~GPU-hours total. Scaling this to Snapshot~2's 1.8M hosts would require roughly an order of magnitude more compute. Beyond cost, the LLM pipeline also requires a separate deterministic refinement stage to suppress recurring false-positive patterns (Section~\ref{sec:labeling}), adding manual effort and engineering complexity.

We therefore train a Random Forest classifier on the Snapshot~1 labels. A tree-based model addresses both limitations: it classifies the full 1.8M-host set in under five minutes on a single CPU, and its learned decision boundaries internalize the suppression logic that the LLM pipeline requires as a separate post-processing stage. The resulting classifier is also interpretable: feature importances identify which signals drive predictions, enabling systematic auditing of classification behavior without probing a black-box model.

\noindent\textbf{Feature Engineering.}
For each host, we extract 25 hand-crafted features from scanner-visible metadata, organized into three groups.

\emph{DER content signals} capture solar- and energy-specific indicators: the number of matched energy-domain terms
(e.g., \texttt{inverter}, \texttt{voltage}, \texttt{yield}), brand-name detection with word-boundary enforcement for short strings (to prevent, e.g., the four-character brand ``Tigo'' from matching ``Tigoni''), solar-related HTTP titles, TLS certificate fields referencing solar platforms, and industrial protocol exposure (SNMP, Modbus).

\emph{Suppressor signals} flag hosts that lack DER evidence.  A \texttt{no\_solar\_content\_signal} feature fires when none of the DER content features are active.  Additional compound features detect many-port hosts with zero DER indicators, lighttpd servers without corroborating solar evidence, and DevOps, AI, or database dashboards (e.g., Jupyter, Grafana, pgAdmin) that are structurally similar to monitoring interfaces but unrelated to energy systems.

\emph{TF-IDF features} provide text-level discrimination using a discriminative vocabulary. \emph{Negative} terms are selected by $\chi^2$~association with the not-solar class, filtered to remove CCTV-specific and web boilerplate, and supplemented with a seeded list of known false-positive indicators (e.g., \texttt{docker}, \texttt{kubernetes}, \texttt{wordpress}). \emph{Positive} terms are drawn from a curated DER phrase list intersected with terms that appear frequently in solar-labeled documents. Remaining vocabulary slots are filled with top-frequency corpus terms as a hybrid backfill.

\noindent\textbf{Training.}
We train a Random Forest (200~trees, max depth~30) on the full set of labeled hosts from Snapshot~1. We limit tree depth to 30 to reduce overfitting to training-set-specific patterns; holdout evaluation confirmed that unrestricted depth did not improve precision or recall. Two hundred trees provide stable probability estimates while keeping training and inference fast enough for iterative development; we observed no improvement beyond this point in cross-validation. No class-weight rebalancing is applied, so that the model's prior reflects the natural class distribution and hosts with no solar signal default to not-solar.

\noindent\textbf{Threshold Selection.}
The Random Forest outputs a probability estimate $P(\text{solar}) \in [0, 1]$ for each host, representing the model's confidence that the host is a DER deployment. We classify a host as solar when $P(\text{solar})$ exceeds a threshold $\tau$; the choice of $\tau$ trades off precision against recall: higher $\tau$ admits fewer but more confident predictions, while lower $\tau$ catches more devices at the cost of false positives. Sweeping $\tau$ from 0.10 to 0.90 (Figure~\ref{fig:threshold_sweep}, Appendix~\ref{app:threshold}) reveals an elbow at $\tau = 0.25$, where the predicted host count transitions from steep decline to gradual decrease. We select this value as our operating point.

\noindent\textbf{Consensus Voting.}
To reduce sensitivity to training randomness, we train five independent models using different random seeds and retain only hosts classified as solar by all five runs (intersection consensus). This further improves precision by excluding hosts whose classification is unstable across training perturbations.

\noindent\textbf{Generalization to New Brands.}
Because Snapshot~2 incorporates 62 vendors not present in Snapshot~1, the classifier must generalize beyond the specific brands it was trained on. It does so through features that capture general DER characteristics (i.e., energy-domain terminology, telemetry patterns, embedded server signatures, and industrial protocol exposure) rather than relying solely on brand-name matching. This design enables the classifier to identify DER hosts from vendors that were absent from the training data, and allows us to assess whether exposure patterns observed in Snapshot~1 persist four months later.

\noindent\textbf{Validation.}
We evaluate the classifier on both the training distribution (Snapshot~1) and the target distribution (Snapshot~2).

On Snapshot~1, five-fold stratified cross-validation yields 98.0\% precision and 86.3\% recall (solar class). A held-out stratified test split confirms these estimates (98.2\% precision, 86.1\% recall; Table~\ref{tab:rf_performance}).

On Snapshot~2, we draw a random sample of 100~predicted-solar and 100~predicted-not-solar hosts and manually inspect each.  Of the 100~predicted-solar hosts, 99 are true DER deployments; the single false positive is a Reolink solar-powered camera that was not represented in the Snapshot~1 training set and evaded the suppression rules developed in Stage~2.  Of the 100~predicted-not-solar hosts, 98 are correctly rejected; the two false negatives are hosts with valid SMA Solar Technology AG TLS certificates and inverter-related metadata, but whose body content lacked sufficient brand or telemetry signal to exceed the classification threshold. This is a consequence of our precision-prioritizing design.

\begin{table}[]
\centering
\small
\caption{Random Forest classification performance on the training distribution (Snapshot~1, ground truth) and the target distribution (Snapshot~2, audit of random samples).}
\label{tab:rf_performance}
\begin{tabular}{@{}llcc@{}}
\toprule
\textbf{Dataset} & \textbf{Evaluation} & \textbf{Precision} & \textbf{Recall} \\
\midrule
Snapshot~1 & Stratified holdout & 98.2\% & 86.1\% \\
Snapshot~2 & Manual audit (n=200) & 99\% & 98\% \\
\bottomrule
\end{tabular}
\label{tab:rf_performance}
\end{table}

The output of this stage is a validated set of 66,379
Internet-exposed solar DER hosts, which serves as the input to
device attribution and security analysis in Stage~4.

\noindent\textbf{\new{Honeypot Handling.}}
\new{Honeypots are a known confound for Internet-exposed ICS/OT measurement and can constitute a non-negligible fraction of such populations. We address them at two levels. First, when constructing the training set, we remove hosts flagged as honeypots by Censys's built-in detection (1,114 hosts, 0.67\%, in Snapshot~1; 50,753, 2.68\%, in Snapshot~2) and reassign them to the not-solar class, instead of discarding them, so the classifier learns to reject honeypot-like hosts even when Censys does not flag them. This is reinforced by our suppressor and negative TF-IDF features, which down-weight generic-infrastructure and non-DER text patterns characteristic of decoys. Crucially, the Censys honeypot flag is not a classifier input feature: on Snapshot~2, the model assigns flagged hosts to the not-solar class on its own. As evidence that residual honeypots do not affect our results, no host in the final validated population of 66,379 carries a Censys honeypot flag, and our manual audits (100 predicted-solar hosts per snapshot) surfaced none. Second, for the 571 devices used in our Oahu power-grid analysis (Section~\ref{sec:power-analysis}), we perform an additional cyber-physical consistency check of the kind described by Tay et al.~\cite{tay2023taxonomy}, manually verifying each against satellite imagery to confirm a real physical installation.}

\subsection{Stage 4: Device and Security Analysis}
\label{sec:security_analysis}

For each validated solar DER host, we extract security-relevant attributes that characterize both its operational role and its exposure to Internet-based attacks. This analysis maps scanner-visible properties to attacker capabilities defined in our threat model and enables population-level measurement of exposed control surfaces.

\noindent\textbf{Vendor and Device Attribution.} We first attribute each host to a vendor and device category using a combination of HTML content, TLS certificate metadata, and extracted model identifiers. Vendor attribution is performed by matching brand names, portal domains, and model numbers against the fingerprint database. Hosts are then classified by device role into inverters, monitoring gateways or loggers, and installer or fleet-management portals based on interface structure and functionality. For example, an SMA host exposing a \emph{Sunny WebBox} interface with per-device telemetry and configuration menus is classified as an inverter or gateway, whereas an interface listing multiple sites under an \emph{account.sma} or \emph{login.sma} portal is classified as an installer or fleet-management portal.

\noindent\textbf{Exposed Interfaces and Misconfigurations.} We then analyze each host for exposed security features and misconfigurations. Authentication controls are assessed by detecting login forms, authentication challenges (e.g., HTTP~401 responses), and unauthenticated access to status or control endpoints. To identify protocol-level exposure, we inspect open services and ports for industrial control protocols such as Modbus~TCP, as well as vendor-specific management interfaces. For instance, some SMA devices expose a web dashboard without authentication while simultaneously advertising an open Modbus service, directly enabling unauthenticated read or control access from the public Internet.

Where web interfaces are accessible, we detect the presence of power and control functionality, including active and reactive power setpoints, disconnect or shutdown controls, reboot actions, and firmware update endpoints. In the SMA example, these controls often appear as documented configuration pages referenced in vendor manuals and exposed verbatim in deployed interfaces. \textit{The presence of such controls indicates the feasibility of curtailment, forced disconnection, or coordinated manipulation attacks}.

\noindent\textbf{Vulnerabilities.} To assess software vulnerabilities, we extract software and firmware version information from HTTP headers, HTML content, and service banners when available. Extracted versions are matched against public vulnerability databases to identify known CVEs affecting exposed components. This includes both DER-specific vulnerabilities and general-purpose issues in embedded web servers and management frameworks used by DER platforms. For example, SMA web interfaces frequently disclose web server versions that can be matched to known vulnerabilities in third-party components. We perform CVE matching only when version identifiers are explicitly present. Because scan-visible version disclosure does not capture configuration, feature enablement, or vendor backports, CVE matches indicate potentially affected deployments rather than confirmed exploitability.

Finally, we associate each host with geographic metadata derived from Censys, including country and autonomous system number (ASN). These attributes enable aggregation of exposure statistics by region, network operator, and vendor, and support the population-level analyses presented in Section~\ref{sec:exposure_results}. The resulting characterization yields a structured per-host record capturing vendor attribution, device role, exposed interfaces, misconfigurations, and known vulnerabilities. The per-host records form the basis for our measurement of real-world DER security posture.

\section{RQ1: Exposure Results}
\label{sec:exposure_results}

We evaluate how solar DER platforms are exposed on the public Internet and what security-relevant control surfaces are reachable in practice. This section addresses \textbf{RQ1}: \emph{How are solar DER platforms exposed on the public Internet, and what security-relevant control surfaces are reachable?}

Using the validated host set produced by our Internet-scale discovery pipeline, we characterize exposure along four dimensions: (i) externally reachable interfaces and protocols, (ii) vendor attribution and deployment heterogeneity, (iii) temporal persistence across snapshots, and (iv) observable software vulnerabilities. Exposure spans over 150 countries; the full geographic distribution is reported in Appendix~\ref{app:geography}.

\subsection{Externally Reachable Interfaces}
\label{sec:externally_reachable}

\begin{table}[t]
\centering
\caption{Externally reachable interfaces observed on Internet-exposed solar DER hosts.}
\resizebox{\columnwidth}{!}{%
\begin{tabular}{llll}
\toprule
\textbf{Interface class} &
\textbf{Service} &
\textbf{\% hosts} &
\textbf{Example capability} \\
\midrule
\multirow{3}{*}{Web (HTTP)}
& Management portals        & 94.7\% & device configuration \\
& Monitoring APIs           & 10.6\% & power telemetry \\
& Fleet portals             & 3.0\%  & multi-site management \\
\midrule
\multirow{3}{*}{Control protocols}
& Modbus/TCP (502)          & 4.8\%  & inverter setpoints \\
& IEC 60870-5-104, DNP3     & 0.7\%  & grid telemetry \\
& Others                    & 0.2\%  & BACnet, etc.
\\
\midrule
\multirow{3}{*}{Remote access}
& SSH                       & 28.9\% & system access \\
& FTP                       & 6.9\%  & firmware files \\
& Telnet / VNC / RDP        & 16.0\% & device consoles \\
\bottomrule
\end{tabular}
}
\label{tab:exposed_interfaces}
\vspace{-1em}
\end{table}

Table~\ref{tab:exposed_interfaces} summarizes the externally reachable interfaces observed across all confirmed solar DER hosts. Interface classes are inferred from scanner-visible evidence, including open ports, service fingerprints, HTTP responses, and protocol metadata extracted from Censys. Each row represents a distinct class of externally reachable functionality; categories are not mutually exclusive.

Web-based interfaces \textit{dominate} the exposed attack surface. Nearly all hosts expose at least one HTTP service, with 95\% presenting web-based management portals that support device configuration actions such as setting operating modes, updating parameters, or triggering maintenance workflows. A smaller but still significant fraction expose monitoring APIs that return live power or status telemetry, and 3\% expose fleet or installer portals that aggregate and manage multiple devices or sites. These fleet interfaces are particularly security-relevant, as compromise enables simultaneous interaction with many inverters. 

Beyond web interfaces, we observe direct exposure of industrial control protocols. Modbus/TCP is reachable on 4.8\% of hosts, typically exposing SunSpec registers that support low-level inverter control. An additional 0.7\% expose grid-oriented protocols such as IEC~60870-5-104 and DNP3, which are designed for trusted operational networks and provide structured telemetry and control semantics. A small fraction of hosts (0.2\%) expose other control protocols, like BACnet and OPCUA. While less common than HTTP, these protocols directly map to inverter state and grid behavior.

Remote access services further expand the exposed surface. Almost 30\% of hosts expose SSH, and a non-trivial fraction expose FTP or remote console services such as Telnet, VNC, or RDP. These services indicate general-purpose system access and increase the likelihood that vulnerabilities in embedded operating systems or credential management translate into full device compromise.

Together, these results show that Internet-exposed solar DERs frequently present interfaces capable of configuration, coordination, or low-level control, rather than being limited to passive monitoring. In fact, in the 2025 distributed cyberattack \cite{cert_pl_2025} on Poland's renewable energy sector, attackers exploited default credentials in the web interfaces and remote-access services (SSH, FTP) to gain control of critical DER devices.

\subsection{Vendor and Deployment Heterogeneity}

\label{sec:hetero}

\begin{table}[t]
\centering
\small
\caption{Distribution of vendor matches per DER host. Hosts matching two or more vendors reflect heterogeneous deployments combining multiple inverters, monitoring gateways, or installer-managed platforms.}
\begin{tabular}{lrr}
\toprule
\textbf{\# vendors per host} & \textbf{Hosts} & \textbf{Share} \\
\midrule
0 vendors (Generic)            & 35,662  & 53.7\% \\
1 vendors             & 25,912 & 39.0\% \\
2 vendors            & 3,726  & 5.6\% \\
3+ vendors           & 1,079  & 1.6\% \\
\midrule
Total               & 66,379 & 100.0\% \\
\bottomrule
\end{tabular}
\label{tab:rq1_brand_matches}
\vspace{-1em}
\end{table}

We next examine how exposed interfaces map to vendors and deployment models. Table~\ref{tab:rq1_brand_matches} reports the number of vendor fingerprints matched per host. Over half of all exposed hosts (53.7\%) cannot be attributed to any of the 96 vendors in our fingerprint set and are labeled as generic. To confirm that these hosts are not simply missed vendor matches, we manually inspected a random sample of 50 generic hosts. In nearly all cases, these interfaces exposed extremely generic language (e.g., ``Solar Inverter v5.0''), minimal branding, and direct access to configuration or control functionality, consistent with installer-specific deployments, white-label platforms, or customized firmware instead of known vendor portals.

Among attributed hosts, most map to a single vendor, but a notable fraction match multiple vendors. These multi-vendor hosts reflect heterogeneous deployments, such as an inverter paired with a third-party monitoring gateway or an installer-managed platform layered on top of vendor firmware. From a security perspective, such layering increases complexity and creates additional opportunities for misconfiguration or inconsistent security controls.

We observe that many generic and multi-vendor deployments expose interfaces that appear to be operated by third-party integrators rather than inverter manufacturers themselves, often providing unified monitoring or management across equipment from multiple vendors. While such integration can simplify operations, it also shifts security responsibility away from manufacturers toward installers or service providers that may not follow consistent security practices. This deployment model helps explain both the prevalence of generic interfaces and their frequent Internet exposure.

\begin{table}[t]
\centering
\small
\caption{Global prevalence of brands across exposed solar DER hosts. Note that a single host may have multiple brands and hosts with no brand match are excluded from this table.}
\begin{tabular}{lrr}
\toprule
\textbf{Vendor} & \textbf{Hosts matched} & \textbf{\% hosts} \\
\midrule
SMA                  &  5,814 &   8.8\% \\
Fronius              &  3,786 &   5.7\% \\
Campbell Scientific  &  3,294 &   5.0\% \\
Auo                  &  2,900 &   4.4\% \\
Huawei               &  2,866 &   4.3\% \\
Tesla Energy         &  1,700 &   2.6\% \\
Chint                &  1,655 &   2.5\% \\
Solarlog             &  1,613 &   2.4\% \\
Tigo                 &  1,486 &   2.2\% \\
REFU                 &  1,165 &   1.8\% \\
Other brands (75)    & 10,492 & 15.8\% \\
\bottomrule
\end{tabular}
\label{tab:rq1_vendor_global}
\end{table}

Table~\ref{tab:rq1_vendor_global} complements this host-level view by showing the global prevalence of vendors across all attributed deployments. Exposure is distributed across a broad set of vendors with a gradual decline in prevalence: SMA appears on 8.8\% of all exposed hosts, followed by Fronius (5.7\%), Campbell Scientific (5.0\%), AUO (4.4\%), and Huawei (4.3\%). Beyond the top ten, each vendor accounts for less than 2\% of hosts, and the bottom 60 vendors each appear on fewer than 0.1\%. Because hosts may match multiple vendors, these percentages are not mutually exclusive.

From a security perspective, this concentration has two implications. First, vulnerabilities or insecure defaults in a small number of widely deployed platforms can affect thousands of Internet-exposed devices simultaneously, amplifying attacker impact. Second, beyond these few dominant vendors, exposure quickly fragments across a long tail of platforms and deployments, reinforcing the challenge of relying on vendor-specific signatures alone: even among branded hosts, no single vendor dominates enough to support comprehensive measurement without accounting for diversity, rebranding, and third-party integration.

\subsection{Persistence and Generalization}

\label{sec:persistence}

To assess whether Internet exposure of solar DERs is persistent and to validate the classifier's ability to generalize beyond its training data, we compare validated DER hosts across the two snapshots taken four months apart.

\noindent\textbf{IP-level persistence.}
Of the 41,755 validated DER hosts in Snapshot~1, 63.7\% (26,605) reappear in Snapshot~2, indicating that the majority of exposed devices remain Internet-facing over a multi-month window. The remaining 36.3\% (15,150) of Snapshot~1 hosts are absent from Snapshot~2, likely due to IP churn, remediation, or changes
in network configuration. Conversely, 59.9\% of Snapshot~2 hosts (39,774) are entirely new, reflecting both the broader query coverage (96 vs.\ 34 vendors) and the natural growth of Internet-exposed DER deployments. This persistence rate implies that exposed DERs constitute a sustained attack surface: an adversary who identifies a vulnerable device has a window of months, not hours, in which to exploit it.

\noindent\textbf{Brand distribution stability.}
To fairly assess whether vendor exposure is stable over time, we restrict comparison to the 34 vendors with queries in both snapshots. Among these shared vendors, the top five are SMA, Fronius, Campbell Scientific, Huawei, and Solar-Log; their rank order is unchanged across both snapshots, and relative shares are consistent (e.g., SMA accounts for 15.2\% of branded hosts in Snapshot~1 and 8.8\% in Snapshot~2, with the decrease reflecting dilution from the expanded vendor set, not reduced SMA exposure). This confirms that the exposure landscape is structurally stable across snapshots, and that the classifier trained on Snapshot~1 labels produces consistent attribution when applied to independent data collected months later.

Examining individual vendors within the shared set reveals movement in both directions. Enphase decreases from 451 to 394 attributed hosts; in correspondence with the authors, Enphase confirmed active efforts to reduce the Internet exposure of their IQ~Gateway devices between the two snapshot periods, suggesting that the observed reduction reflects deliberate remediation. Contec, whose SolarView Compact monitors were targeted by Mirai-based botnets in 2024~\cite{dersec_2025}, also decreases from 686 to 512 hosts, possibly reflecting operator response to publicized exploitation. Solis shows the steepest relative decline among shared vendors, dropping from 488 to 193 hosts. In the other direction, Growatt increases from 939 to 1,128 hosts, consistent with Growatt's rapid global market expansion as the world's largest residential inverter supplier. These per-vendor trends indicate that while the overall exposure landscape is stable, individual vendors' footprints do shift: in some cases due to active security remediation, and in others due to continued deployment growth outpacing any hardening efforts.

\noindent\textbf{Generalization to unseen vendors.}
Snapshot~2 incorporates 62 vendors absent from the Snapshot~1 query set and therefore absent from the training data. The classifier successfully identifies hosts for 6 of these previously unseen brands (46~hosts), including Also Energy, Span, Skyfri, and Omnidian. While the absolute count is modest, these detections confirm that the feature set, built around energy-domain terminology, telemetry patterns, and embedded server signatures, generalizes beyond brand-specific memorization. The remaining 56 new-query vendors yield zero attributed hosts after classification. While vendor-specific queries for these brands do return Censys results, the returned hosts consist of marketing pages, documentation sites, and other non-device infrastructure that the classifier correctly rejects.

\noindent\textbf{Security consequences.}
Together, these results establish that Internet exposure of solar DER platforms is both persistent and structurally stable. The high IP overlap rate means that vulnerability windows extend across months, giving adversaries sustained access to exposed control surfaces. The stable vendor distribution across independent snapshots confirms that exposure arises from systemic deployment practices rather than transient misconfigurations, and that effective mitigation will require changes to how DERs are installed and managed at scale.

\subsection{Observed Software Vulnerabilities}

\label{sec:vulns}

\begin{table}[t]
\centering
\small
\caption{Potential CVEs observed on Internet-exposed solar DER hosts grouped by brand or device component.}
\begin{tabular}{lrrr}
\toprule
\textbf{Vulnerable component} &
\textbf{\# CVEs} &
\textbf{Affected hosts} &
\textbf{\% hosts} \\
\midrule
SMA $^\diamond$IF                    & 16 &  4,738 &  7.1\% \\
Fronius $^\diamond$IF                &  2 &    2,175 &   3.3\% \\
Growatt $^\diamond$IF                &  10 &    889 &   1.3\% \\
Contec $^\ddagger$SC           & 10 &    486 &   0.7\% \\
Solar-Log $^\diamond$IF              & 10 &    328 &   0.5\% \\
Enphase $^\diamond$IF                & 10 &    234 &   0.4\% \\
SolarEdge $^\bullet$MI           &  2 &    71 &   0.1\% \\
Sungrow $^\diamond$IF                &  10 &     54 &   0.1\% \\
APsystems $^\diamond$IF           &  4 &      17 &   0.0\% \\
Other DER devices &  26 &     2,834 &   4.3\% \\
\midrule
\textbf{Total}                           & 100 &  11,826 & -- \\
\bottomrule
\end{tabular}\\
\label{tab:rq2_cve_summary}
{$^\diamond$IF = inverter firmware; $^\bullet$MI = management interface; \\$^\ddagger$SC = SolarView Compact}
\vspace{-1em}
\end{table}

Finally, we examine known software vulnerabilities visible on exposed hosts. Table~\ref{tab:rq2_cve_summary} aggregates observed CVEs by affected component, drawing on publicly reported DER-related vulnerabilities~\cite{dersec_2025}. CVEs are identified by extracting explicit firmware or software version identifiers from banners, HTTP responses, and protocol metadata, and matching them against public vulnerability databases.

Vulnerabilities are highly concentrated within specific vendor firmware and management components, reflecting the underlying vendor distribution of exposed devices. In particular, SMA inverter firmware accounts for almost 5,000 affected hosts across 16 distinct CVEs, consistent with SMA being the most prevalent branded platform in our dataset (Table~\ref{tab:rq1_vendor_global}). Several other widely deployed vendors also exhibit smaller but non-trivial vulnerable populations.

These vulnerabilities frequently affect web management interfaces or embedded services that are directly reachable from the public Internet. As a result, a meaningful fraction of exposed DERs run software with known, documented weaknesses. We note that CVE presence indicates potential vulnerability based on version matching, not confirmed exploitability; however, when combined with the externally reachable control surfaces shown earlier, these vulnerabilities substantially increase the feasibility of remote compromise at scale.

Taken together, our results for RQ1 show that Internet-exposed solar DER platforms frequently expose management, coordination, and control interfaces; that exposure spans a fragmented and partially unbranded ecosystem; and that many exposed devices run potentially vulnerable software. This combination of reachable control surfaces and known weaknesses motivates the grid-level impact analysis that follows, where we examine what an attacker could achieve by exploiting this exposure in a realistic power system.

Finally, we use the results of this pipeline to ground our power-system impact analysis. From the globally identified host set, we isolate all exposed solar DERs located in Oahu, Hawaii. Each of these hosts is then manually inspected to confirm the presence of severe, attacker-relevant exposures, including unauthenticated management interfaces, firmware upload endpoints, or externally reachable control protocols such as Modbus/TCP. Only devices that satisfy this conservative validation are included in the subsequent power grid analysis. This ensures that the grid-level results presented in the next section are based exclusively on real, Internet-exposed inverters with demonstrable control or disruption potential.

\section{RQ2: Power Grid Analysis}
\label{sec:power-analysis}
This section demonstrates the effects of the attacker's efforts on a realistic power grid model. An attacker may have multiple objectives when compromising DERs, such as destroying individual DER devices, disrupting network stability, and affecting operational aspects such as electricity markets~\cite{bhaskar_threat_2024,xiangyu_destabilizing_2025}. This section analyzes a scenario in which the attacker causes significant violations of steady-state voltage and current flow limits. Specifically, we seek to determine the worst possible disruptions to bus voltages and transmission line flows that an attacker can cause with full control of all compromisable solar DERs. We analyze this using the DERs identified in Section~\ref{sec:exposure_results} along with a power grid model of the Hawaiian island of Oahu.

We select Oahu for this study since it is on the leading edge of the broader trend of accelerating solar DER deployments. DER penetration in Oahu is higher than most other regions in the country~\cite{nrel_solar_update_2019}, making Oahu especially valuable for  studying grid-level risks as DER deployments continue to grow. This is consistent with many power system studies that use the Oahu power grid as a test region for high-DER integration, e.g.,~\cite{nrel_pv_oahu_2013, nrel_owits_2011}.

\subsection{Testbed Description}
\begin{figure}[]
    \centering
    \includegraphics[width=0.99\linewidth]{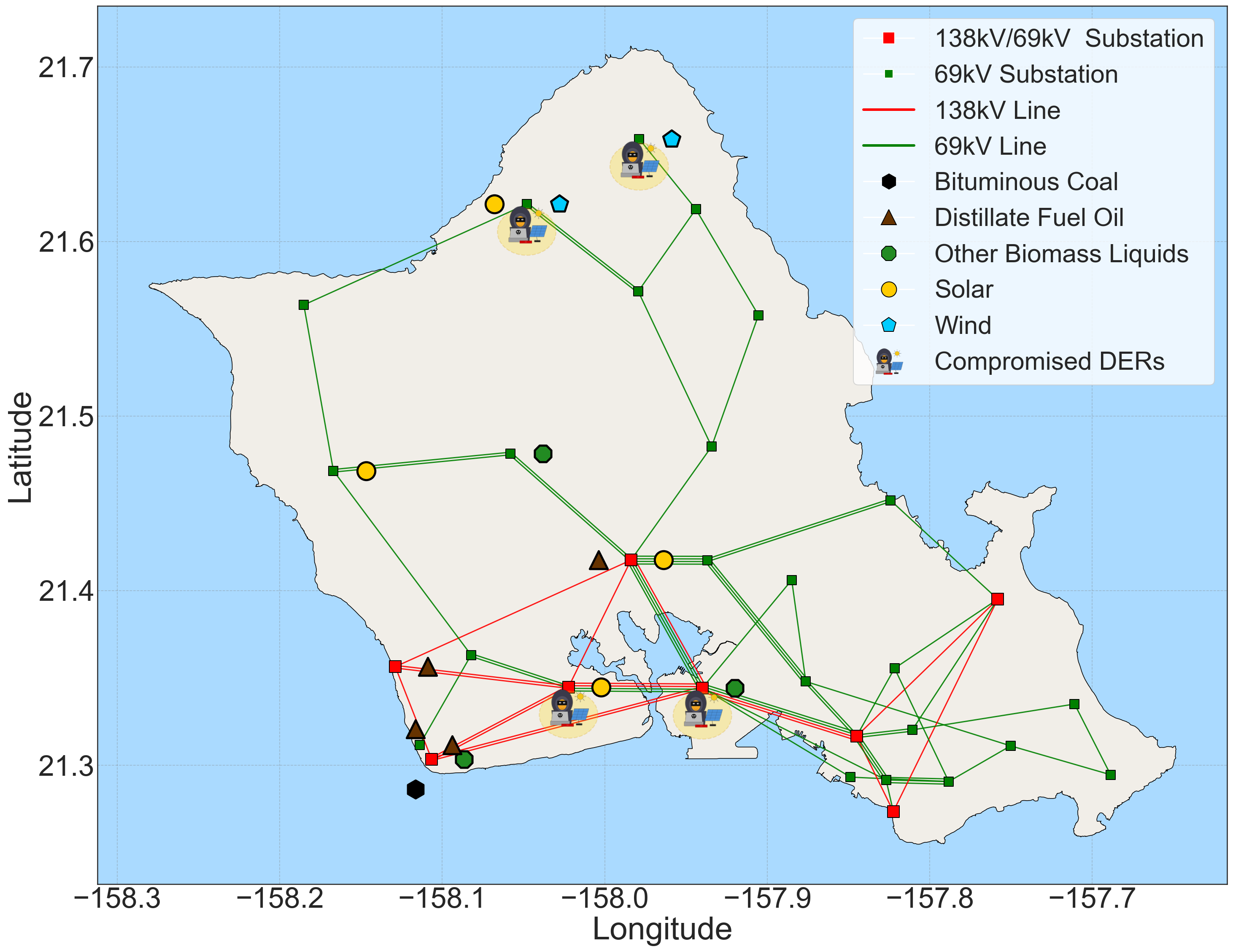}
    \caption{A diagram of the synthetic electricity grid network of the Oahu island, Hawaii~\cite{tamu_electric_grids_website}.}
    \label{fig:oahu}
    \vspace{-1em}
\end{figure}

Our analysis uses a power grid model of Oahu from~\cite{tamu_electric_grids_website}, developed and validated as part of the US Department of Energy's GRID DATA program~\cite{grid_data} and widely used by the power systems research community. This model is comprised of a 37-bus electric transmission system covering Oahu's geographic footprint.
As part of the GRID DATA program, this model and all other models provided in~\cite{tamu_electric_grids_website} have been validated for realism and accuracy~\cite{grid_structural_characteristics}. The system serves a peak load of 1100 MW. To represent realistic reactive power demands, we set load power factors of $0.95$ lagging. The grid model is visualized in Figure~\ref{fig:oahu}. 

Leveraging the results from Section~\ref{sec:exposure_results}, we augment the original Oahu system from~\cite{tamu_electric_grids_website} with the compromisable solar DERs we identified in Oahu. We manually verified each compromisable DER's location via Google Maps satellite imagery, confirming visible solar panels at the expected sites\new{, thus providing the cyber-physical consistency check noted in Section~\ref{sec:scaled_classification} and ensuring the grid analysis rests on real installations and not decoy honeypots}. Consistent with typical transmission system modeling practices, we represent the distribution networks connected to the transmission system by equivalent power injections at each bus. Accordingly, we attach each compromisable solar DER model to the geographically closest transmission system bus. As shown in Table~\ref{tab:grid-augmentation}, our augmented test case contains the original system's 41 generators, 6 of which are solar units, with 41.1~MW of additional compromisable solar DERs.

\begin{table}[]
    \centering
    \caption{Power grid model augmented with the identified exposed inverters. The original case refers to the Oahu network from~\cite{tamu_electric_grids_website}. The augmented case adds compromisable solar DERs in Oahu as identified in Section~\ref{sec:exposure_results}.}

    \begin{tabular}{lcc}
         & Original  & Augmented  \\
        \midrule
        Total solar capacity (MW) & $167.4$ & $208.5$ \\
        Compromisable capacity (MW) & $\phantom{00}0\phantom{.0}$ & $\phantom{0}41.1$ \\

        \bottomrule
    \end{tabular}
    \label{tab:grid-augmentation}
\end{table}

\subsection{Problem Formulation}
To model the attacker's actions upon controlling all of the compromisible DERs, we formulate and solve optimization problems that manipulate the DERs' active and reactive power injections to maximize disruptions to the voltage magnitudes and line flows. \new{This yields a worst-case upper bound on physically achievable impact: it characterizes what the AC power-flow physics permit an attacker controlling the compromised inverters to do, independent of assumptions about that attacker's knowledge or coordination. The device exposure underlying this set is empirically validated with each Oahu DER manually confirmed as severely exposed; conversely, full simultaneous control is an assumption we make to bound the worst case.} We next formalize these optimization problems by modeling the transmission network, both compromised and non-compromised generators, and the attacker's objectives.

\noindent\textbf{Network and Component Models.}
We consider a power network defined by buses~$\mathcal{N}$, lines~$\mathcal{E}\subseteq \mathcal{N}\times\mathcal{N}$, and generators~$\mathcal{G}$. 
The current flow through line~$(i,j)\in\mathcal{E}$ is denoted as $I_{ij}$, with $I_{ij}^{ft}$ and $I_{ij}^{tf}$ indicating flows into the bus~$i$ terminal (``from bus'') and the bus~$j$ terminal (``to bus''), respectively. Each bus~$i\in\mathcal{N}$ has a voltage phasor $V_i\angle \theta_i$ as well as active and reactive power demands~$P_i^{d}$ and $Q_i^{d}$. Line flows and voltages have limits required for acceptable system operation: $|I_{ij}^{ft}|\le I_{ij}^{\max}$, $|I_{ij}^{tf}|\le I_{ij}^{\max}$, and $V_i^{\min}\le |V_i|\le V_i^{\max}$.

\begin{figure}[!t]
\centering
\begin{tikzpicture}[scale=2]
    \draw[-{Stealth}] (-0.2,0) -- (1.4,0) node[right] {$P_k$};
    \draw[-{Stealth}] (0,-1.2) -- (0,1.2) node[above] {$Q_k$};

    \fill[red!15, opacity=0.5] (0,0) -- (0,-1) arc (-90:90:1) -- cycle;

    \draw[thick] (0,-1) arc (-90:90:1);
    \draw[thick] (0,-1) -- (0,1);  %

    \draw[densely dotted, gray] (0,0) -- (0.707,0.707);
    \node[gray, font=\footnotesize, anchor=south east] at (1,0.65) {$S_k^{\max}$};

    \fill[green!60!black] (0.85,0.1) circle (1.5pt);
    \node[green!60!black, anchor=west, font=\footnotesize] at (0.35,0.1) {normal};

    \fill[red] (0.436,0.9) circle (1.5pt);
    \node[red, anchor=south, font=\footnotesize] at (0.436,0.93) {attack};

    \draw[-{Stealth}, red, thick, densely dashed] (0.80,0.14) to[bend left=35] (0.44,0.85);

    \node[font=\small] at (0.6,-1.15) {$P_k^2 + Q_k^2 \le (S_k^{\max})^2$};
\end{tikzpicture}
\caption{Operating region for inverter $k\in\mathcal{G}_c$. The red area is the attacker's feasible region bounded by the inverter's apparent power rating $S_k^{\max}$ (arc). The attacker can change the inverter's output from the normal operating point (green) to a malicious one, e.g., one with large reactive power injection (red), to induce violations of voltage and flow limits.}
\label{fig:pq-circle}
\vspace{-1em}
\end{figure}

\noindent\textbf{Generator Modeling.}
The set of generators $\mathcal{G}$ is partitioned into non-compromised generators in the set $\mathcal{G}_{nc} \subset \mathcal{G}$ and inverters corresponding to compromised DERs in the set $\mathcal{G}_c \subset \mathcal{G}$. The compromised inverters $k\in\mathcal{G}_c$ are constrained by their apparent power rating~$S_k^{\max}$ and a lower bound on active power output $P_k^{\min} = 0$ corresponding to curtailable solar production (cf Figure~\ref{fig:pq-circle}). The attacker controls the compromised inverters' active and reactive outputs within this operating region.

Following typical power system modeling practices, non-\hspace{0pt}compromised generators respond automatically to system conditions. For active power, generators participate in \emph{automatic generation control} (AGC) via droop response modeled with a distributed slack bus formulation~\cite{Dhople2020DistributedSlack}: ideally, $P_k = P_k^0 + \alpha_k \Delta$ for each generator $k\in\mathcal{G}_{nc}$, where $P_k^0$ is the nominal setpoint, $\alpha_k$ is the specified participation factor, and $\Delta$ is a system-wide frequency deviation signal. This linear relationship saturates at the generator's physical limits $P_k^{\min}$ and $P_k^{\max}$. We model this with a smooth sigmoid approximation:
\begin{equation}\label{eq:fagc}
f_{\text{AGC}}(\Delta) = P_k^{\min} + \frac{P_k^{\max} - P_k^{\min}}{1 + e^{-\kappa_k(\Delta - \Delta^0_{k})}},
\end{equation}
where $\kappa_k$ controls the steepness and $\Delta^0_{k} = \kappa_k^{-1}\ln\bigl((P_k^{\max} - P_k^0)/(P_k^0 - P_k^{\min})\bigr)$ ensures the curve passes through the nominal operating point ($P_k^0$ when $\Delta = 0$). The parameter $\kappa_k$ is chosen to best approximate the piecewise linear droop curve.

Each generator $k\in\mathcal{G}_{nc}$ in PV mode regulates its terminal voltage to its setpoint $V_k^*$ by adjusting its reactive output $Q_k$. When $Q_k$ reaches its upper or lower limit ($Q_k^{\max}$ or $Q_k^{\min}$, respectively), the generator switches to PQ mode (constant reactive power). We model this piecewise linear \emph{PV/PQ switching} characteristic~\cite{zhao2008bus} with a smooth sigmoid approximation:
\begin{equation}\label{eq:fpvpq}
f_{\text{PV/PQ}}(V) = Q_k^{\min} + \frac{Q_k^{\max} - Q_k^{\min}}{1 + e^{\kappa_{\text{pv}}(V - V_k^*) + \ln\bigl(\frac{Q_k^{\max} - Q_k^0}{Q_k^0 - Q_k^{\min}}\bigr)}},
\end{equation}
where $\kappa_{\text{pv}}$ controls the switching sharpness and the logarithmic term ensures the curve passes through the nominal point $(V_k^*, Q_k^0)$. Both sigmoid functions are illustrated in Figure~\ref{fig:scurve-control}.

\noindent\textbf{Worst-case attack model.}
The attacker computes the \emph{worst possible} attack via adversarial AC optimal power flow (AC OPF) optimization problems. To identify the worst-case violations of operational requirements, the attacker solves separate optimization problems targeting distinct failure modes:
\begin{enumerate}[nosep,leftmargin=*]
    \item \textbf{Line overloading (from$\to$to direction)}: $\max |I_{ij}^{ft}|^2$ for target line $(i,j)\in\mathcal{E}$.
    \item \textbf{Line overloading (to$\to$from direction)}: $\max |I_{ij}^{tf}|^2$ for target line $(i,j)\in\mathcal{E}$.
    \item \textbf{Overvoltage}: $\max |V_i|$ for target bus $i\in\mathcal{N}$.
    \item \textbf{Undervoltage}: $\min |V_i|$ for target bus $i\in\mathcal{N}$.
\end{enumerate}
Each optimization problem targets a single component, yielding $2|\mathcal{E}| + 2|\mathcal{N}|$ independent problems, each with different objective functions but sharing the same constraints. For line currents, we report $\max(|I_{ij}^{ft}|, |I_{ij}^{tf}|)$ for each line. 
Conceptually similar to the subproblems in robust OPF algorithms that are used to manage uncertain renewable generation~\cite{acopf_robust2018,roald_pscc2022_survey}, the general form of each attack optimization is:
\begin{subequations}\label{eq:attack-opt}
\begin{align}
\max_{P,Q,V,\theta,\Delta} \quad & f_{\text{attack}}(\cdot) \label{eq:obj}\\
\text{s.t.} \quad & P_i^g - P_i^d = \sum_{j:(i,j)\in\mathcal{E}} P_{ij}(V,\theta) & \forall i\in \mathcal{N}\label{eq:pbal}\\
& Q_i^g - Q_i^d = \sum_{j:(i,j)\in\mathcal{E}} Q_{ij}(V,\theta) & \forall i\in \mathcal{N} \label{eq:qbal}\\
& P_k^2 + Q_k^2 \le (S_k^{\max})^2 & \forall k \in \mathcal{G}_c \label{eq:smax}\\
& 0 \le P_k \le P_k^{\max} & \forall k \in \mathcal{G}_c \label{eq:pbound}\\
& P_k = f_{\text{AGC}}(\Delta; P_k^0, \alpha_k) & \forall k \in \mathcal{G}_{nc} \label{eq:agc}\\
& Q_k = f_{\text{PV/PQ}}(V_{i(k)}; Q_k^0, V_k^*) &  \forall k \in \mathcal{G}_{nc} \label{eq:pvpq}
\end{align}
\end{subequations}
where $f_{\text{attack}}$ is $|I_{ij}^{ft}|$, $|I_{ij}^{tf}|$, $|V_i|$, or $-|V_i|$ based on the attack objective. Constraints~\eqref{eq:pbal}--\eqref{eq:qbal} enforce AC power balance at each bus, where $P_{ij}(V,\theta)$ and $Q_{ij}(V,\theta)$ are the standard AC power flow equations~\cite{molzahn_hiskens-fnt2019} and $P^g_i$ and $Q^g_i$ denote the total power generation at bus~$i\in\mathcal{N}$ from all generators at that bus.

For compromised inverters, constraints~\eqref{eq:smax}--\eqref{eq:pbound} bound the apparent power injection to the inverter's rated capacity---this half-circle region in the P-Q plane represents the physical limit of the power electronics; see Figure~\ref{fig:pq-circle}. 

For non-compromised generators, \eqref{eq:agc}--\eqref{eq:pvpq} enforce the sigmoid control responses in~\eqref{eq:fagc}--\eqref{eq:fpvpq}. These sigmoid approximations ensure differentiability for numerical optimization while capturing realistic generator characteristics~\cite{kundur_power_1994,Dhople2020DistributedSlack,zhao2008bus}. The nominal setpoints $P_k^0$, $Q_k^0$, and $V_k^\ast$ are obtained via a cost-minimizing AC OPF problem which ensures that all nominal voltages and line flows are within limits. 

Our implementation of~\eqref{eq:attack-opt} uses customized code in \mbox{MATLAB} with the \mbox{FMINCON} nonlinear solver along with the modeling tools \mbox{YALMIP}~\cite{yalmip} and \mbox{M{\sc atpower}}~\cite{matpower}.

\begin{figure}[]
\centering
\begin{tikzpicture}[scale=0.85]
    \begin{scope}[xshift=0cm]
        \draw[-{Stealth}] (-1.8,0) -- (1.8,0) node[right, font=\footnotesize] {$\Delta$};
        \draw[-{Stealth}] (0,-1.5) -- (0,1.8) node[right, font=\footnotesize] {$P_k$};

        \draw[gray, dashed] (-1.5,-1.5) -- (1.5,1.5);

        \draw[thick, blue] plot[smooth, domain=-1.5:1.5, samples=50]
            (\x, {1.2*tanh(1.5*\x)});

        \draw[densely dotted, gray] (-1.5,1.2) -- (1.5,1.2);
        \draw[densely dotted, gray] (-1.5,-1.2) -- (1.5,-1.2);

        \node[anchor=west, font=\scriptsize] at (1.55,1.2) {$P_k^{\max}$};
        \node[anchor=west, font=\scriptsize] at (1.55,-1.2) {$P_k^{\min}$};

        \fill[green!60!black] (0,0) circle (2pt);
        \node[green!60!black, anchor=south west, font=\scriptsize] at (0.05,0.05) {$P_k^0$};

        \node[font=\footnotesize, anchor=south] at (0,2.0) {(a) $f_{\text{AGC}}(\Delta)$};
    \end{scope}

    \begin{scope}[xshift=4.5cm]
        \draw[-{Stealth}] (-1.8,0) -- (1.8,0) node[right, font=\footnotesize] {$V - V_k^*$};
        \draw[-{Stealth}] (0,-1.5) -- (0,1.8) node[right, font=\footnotesize] {$Q_k$};

        \draw[thick, red!70!black] plot[smooth, domain=-1.5:1.5, samples=100]
            (\x, {-1.2 + 2.4/(1 + exp(6.25*\x))});

        \draw[densely dotted, gray] (-1.5,1.2) -- (1.5,1.2);
        \draw[densely dotted, gray] (-1.5,-1.2) -- (1.5,-1.2);

        \node[anchor=west, font=\scriptsize] at (1.55,1.2) {$Q_k^{\max}$};
        \node[anchor=west, font=\scriptsize] at (1.55,-1.2) {$Q_k^{\min}$};

        \fill[green!60!black] (0,0) circle (2pt);
        \node[green!60!black, anchor=south west, font=\scriptsize] at (0.05,0.05) {$Q_k^0$};

        \node[font=\footnotesize, anchor=south] at (0,2.0) {(b) $f_{\text{PV/PQ}}(V)$};
    \end{scope}
\end{tikzpicture}
\caption{Sigmoid responses for non-compromised generators. (a)~AGC: active power $P_k$ follows a smoothed droop response saturating at $P_k^{\min}$ and $P_k^{\max}$; the dashed line is the idealized linear droop $P_k = P_k^0 + \alpha_k\Delta$. (b)~PV/PQ switching: within limits $Q_k^{\min}$ and $Q_k^{\max}$, the reactive power $Q_k$ adjusts to regulate voltage.}%
\label{fig:scurve-control}
\vspace{-1em}
\end{figure}

\subsection{\new{Results}}\label{sec:res}
\begin{figure}[]
    \centering
    \includegraphics[width=0.95\linewidth]{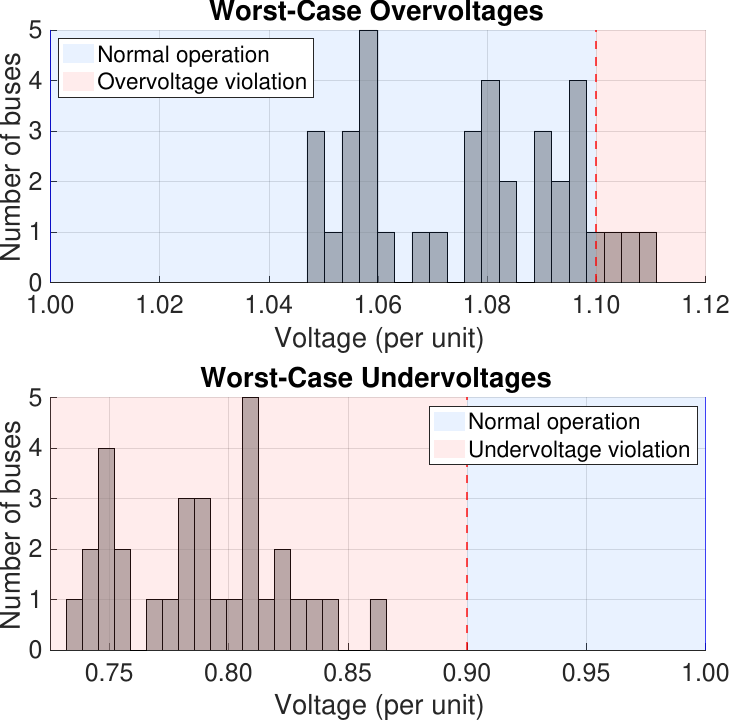} %
    \caption{Histograms of worst-case bus voltage magnitudes across attack scenarios. The figure shows two plots in which the attacker's goal is to cause an overvoltage (upper plot) or an undervoltage (lower plot). The red dashed line indicates the acceptable voltage range (0.90--1.10 per unit) per the Oahu test case's specifications~\cite{tamu_electric_grids_website}, consistent with power quality standards such as ANSI C84.1~\cite{ansi_c84_1}, IEEE 1547~\cite{photovoltaics2018ieee}, and NEMA MG1~\cite{nema_MG1}. The lower voltage limits can be violated significantly, potentially causing outages and equipment damage.}
    \label{fig:voltage-results}
    
    \vspace{-1em}
\end{figure}

\begin{figure}[]
    \centering
    \includegraphics[width=0.95\linewidth]{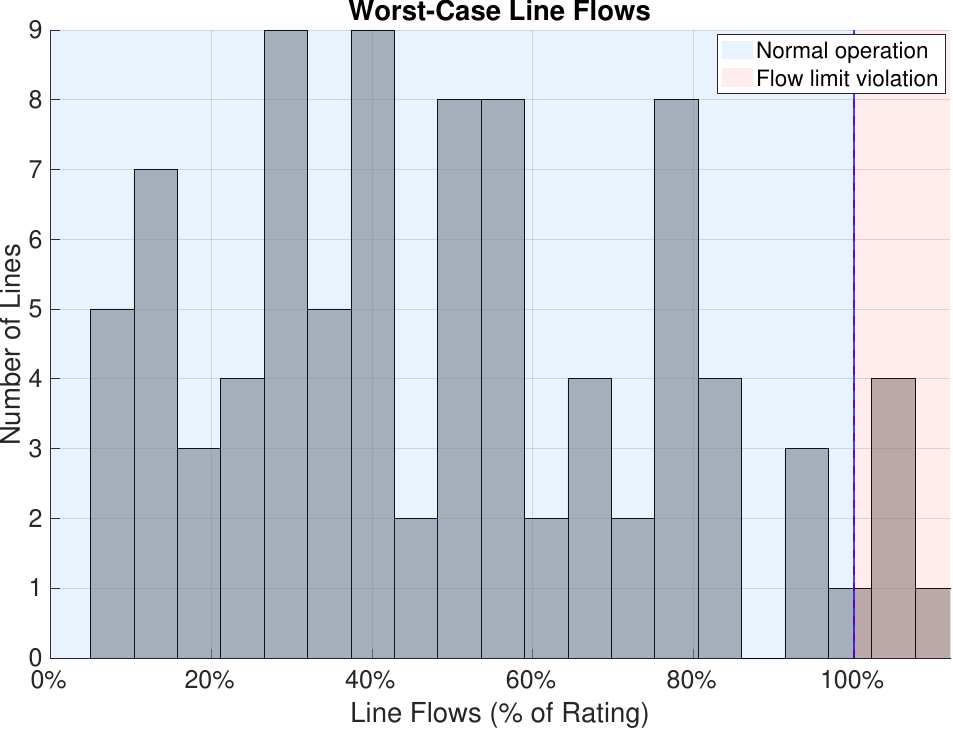} %
    \caption{Histogram of the worst-case line flow limit violations. The nominal rating for each line is at 100\%. The attacker can overload \new{6} out of 89 lines.}
    \label{fig:flow-results}
    \vspace{-1em}
\end{figure}

Solving~\eqref{eq:attack-opt} for each line and bus in the Oahu system yields the worst-case attacks achievable by maliciously controlling the compromised DERs. Figures~\ref{fig:voltage-results} and \ref{fig:flow-results} show the Oahu grid's vulnerability to overvoltages, undervoltages, and line flow violations, with undervoltages being particularly significant.

For each bus~$i\in\mathcal{N}$, the voltage limits given in~\cite{tamu_electric_grids_website} for the Oahu test case are $V_i^{\min} = 0.90$~per unit and $V_i^{\max} = 1.10$~per unit, i.e., $\pm10$\% of the nominal voltage. These limits are consistent with power quality standards such as ANSI C84.1~\cite{ansi_c84_1}, IEEE 1547~\cite{photovoltaics2018ieee}, and NEMA MG1~\cite{nema_MG1}.  As shown by the upper plot in Figure~\ref{fig:voltage-results}, the attacker can force overvoltages beyond 1.10 per unit at 4 of the 37 buses (10\% of the system), albeit by relatively small amounts (maximum voltages of 1.11 per unit). Overvoltages elevate dielectric stress, which can compromise insulation integrity and increase long-term failure probabilities.

More strikingly, the lower plot in Figure~\ref{fig:voltage-results} shows that the attacker can cause significant undervoltages at \textit{any bus in the Oahu network}. Nearly half of the buses in the system (18 out of 37 buses) have worst-case voltages below 0.80 per unit, with some as low as 0.73 per unit. Compared with the lower voltage limits of 0.90 per unit, these undervoltages are far below the acceptable range for this test case. Undervoltage conditions necessitate increased current draw for constant power loads, elevating resistive losses and inducing substantial thermal stresses on conductors, transformers, and motors. Consuming over half of all electric power in aggregate~\cite{iea_motors}, motors are particularly sensitive to undervoltages. Undervoltages of this severity could cause substantial motor malfunctions and significantly reduce lifespans~\cite{nema_MG1}.

The worst-case line flows in Figure~\ref{fig:flow-results} are also concerning, with \new{6} out of 89 lines (6.7\% of the system) being overloaded, sometimes up to 12\% beyond their nominal ratings. This is near the typical range of emergency ratings where corrective actions must be taken within a few minutes to avoid significant overheating and increased sagging~\cite{emergency_ratings}. Operators may have to disconnect these lines to avoid equipment damage, possibly leading to power outages. Simultaneously overloading multiple lines could cause system-wide blackouts.

In summary, the results show that \textit{even with today's levels of exposed inverters, attacks could cause significant power grid disruptions}. As DER penetration increases, future scenarios could have even greater potential for outages and component failures. We next discuss some nuances and implications of these results.

\begin{figure}[]
    \centering
    \includegraphics[width=0.95\linewidth]{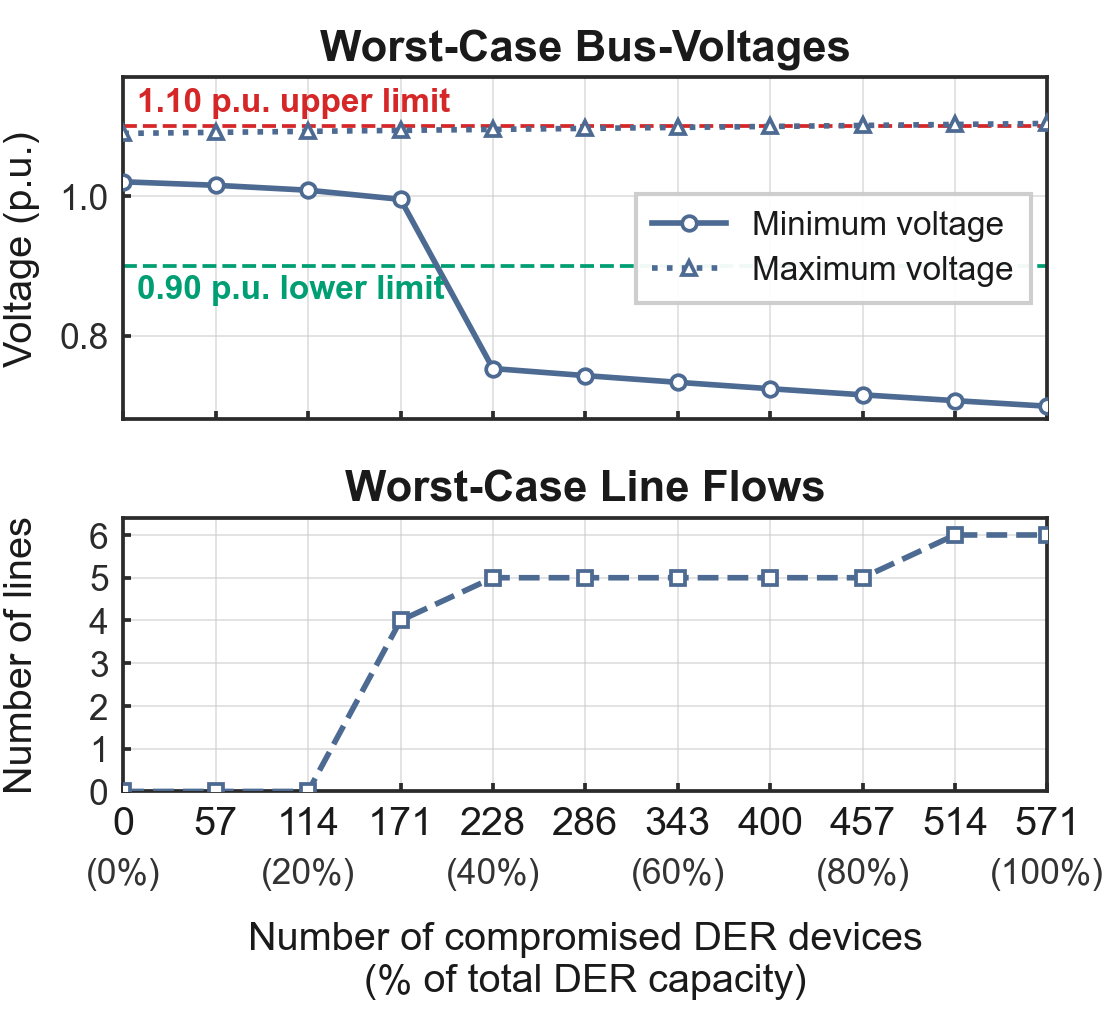} 
    \caption{\new{Worst-case grid impacts in Oahu as the number of compromised DER devices increases. The upper plot shows the minimum and maximum attainable bus voltages, with the green and red dashed lines indicating the 0.90 and 1.10 p.u. limits, respectively. The lower plot shows the number of individually overloadable lines. The horizontal axis reports both the equivalent number of compromised devices and the corresponding percentage of total DER capacity. Across compromise levels evaluated in 10 percentage-point increments, undervoltage first appears at 40\% compromise, while four lines become overloadable at 30\%, five at 40\%, and six at 90\%.}}
    \label{fig:sensitivity}
    \vspace{-1em}
\end{figure}

\noindent\textbf{\new{Sensitivity to the Compromised Fraction.}}
\new{The preceding results assume the attacker controls the full exposed population. In practice, an adversary may control only a subset, and defenses act precisely by shrinking that subset. Figure~\ref{fig:sensitivity} therefore evaluates worst-case grid impact as the compromised fraction varies from 0\% to 100\% in increments of 10 percentage points, reported both as a fraction of total DER capacity and as the equivalent number of compromised devices. Impact scales smoothly with the compromised fraction: undervoltage violations first appear at 40\% compromise, and the number of individually overloadable lines grows from four at 30\%, to five at 40\%, to six at 90\%.

This curve doubles as a quantitative measure of defense effectiveness. The mitigations we discuss in Section~\ref{sec:discussion}, restricting public exposure, requiring VPN or allowlisted access, and strengthening authentication, all reduce the set of devices an attacker can reach. Reading the analysis in reverse, driving the reachable population below 40\% of capacity eliminates undervoltage violations entirely, and smaller reductions yield proportionally fewer overloaded lines. Practical, deployment-level defenses thus translate directly into measurable reductions in grid risk.}

\noindent\textbf{Interpretation of Worst-Case Analysis.}
The optimization formulation~\eqref{eq:attack-opt} computes an upper bound on attack impact---the worst outcome given the attacker's control capabilities. \new{The exposure of these specific devices is established empirically; the attacker's knowledge, coordination, and simultaneous control of them are assumptions that upper-bound the achievable impact.} Real attacks may achieve less impact due to imperfect attacker knowledge of network topology, measurement noise, or defensive responses. We also note that the worst-case impacts are not simultaneously achievable, as the attack that maximizes the impact on one component does not generally yield the worst-case impact for another component. However, the worst-case analysis reveals the fundamental vulnerabilities: the physics of power flow permit adversarial manipulation to cause significant violations of voltage and line flow limits.

\section{Concluding Discussion}

\label{sec:discussion}

Power systems are undergoing a structural shift from centralized generation to grid-connected assets deployed in homes and small commercial settings. These DERs rely on Internet-connected interfaces for monitoring, configuration, and fleet management, placing operational control surfaces outside traditional utility security boundaries. As a result, devices deployed for convenience and scalability can directly influence grid behavior through publicly reachable interfaces. This risk is not hypothetical: the December 2025 cyberattack on Poland’s renewable energy sector demonstrated how exposed DERs can be exploited in practice.

Grounded in this context, our Internet-scale measurement shows that exposed solar DERs form a global, control-capable attack surface. Across 66,379 validated hosts, most expose web-based management portals enabling direct configuration, while smaller but operationally significant fractions expose fleet portals, remote-access services, and industrial control protocols (Table~\ref{tab:exposed_interfaces}). When this empirically measured exposure is incorporated into a power grid model, coordinated manipulation of 41.1~MW of exposed inverters in Oahu drives 18 of 37 buses more than 10\% below nominal voltage and overloads \new{6} of 89 transmission lines (Table~\ref{tab:grid-augmentation}, Figures~\ref{fig:voltage-results}, \ref{fig:flow-results}) -- conditions that can damage equipment and force protective disconnections.

\paragraph{\textbf{Deployment-Driven Exposure.}}
A striking signal in our measurement is the prevalence of generic and layered deployments. Over half of exposed hosts present unbranded interfaces, and a non-trivial subset match multiple vendors, consistent with inverter-gateway stacks and installer-managed dashboards (Table~\ref{tab:rq1_brand_matches}). Even with secure manufacturer defaults, exposure is introduced downstream through installer choices, enabling remote access, forwarding management ports, or aggregating multiple vendors behind one interface. In practice, the effective security boundary is often defined by integrators rather than manufacturers, suggesting that improving DER security requires addressing installer and third-party operational practices alongside firmware vulnerabilities (Table~\ref{tab:rq2_cve_summary}).

\paragraph{\textbf{Persistent Exposure as a Sustained Attack Surface.}} Comparing snapshots taken four months apart (Section~\ref{sec:persistence}), 63.7\% of Snapshot~1 hosts reappear in Snapshot~2, with structurally stable vendor distributions. Adversaries who identify a vulnerable device therefore have a window of months, \textit{not} hours, to exploit it. The same stability makes remediation visible: Enphase confirmed active attempts to reduce IQ Gateway exposure between snapshots, and Contec, whose SolarView monitors were targeted by Mirai botnets in 2024~\cite{dersec_2025}, also declined. Exposure is durable enough to enable sustained access but can be reduced through deliberate action.

\paragraph{\textbf{Implications and Mitigations.}}
Our measurements suggest concrete mitigations across stakeholders, targeting the interfaces that dominate exposure (Table~\ref{tab:exposed_interfaces}):
\new{\begin{itemize} [leftmargin=*,nosep]
\item \textbf{Vendors:} ship secure-by-default exposure modes, stronger authentication, and role separation on management and fleet portals, which dominate the exposed attack surface (Section~\ref{sec:externally_reachable}, Table~\ref{tab:exposed_interfaces}). Fleet and installer portals warrant priority, as a single compromise scales across many inverters.
\item \textbf{Installers/operators:} keep control protocols and remote-access services off the public Internet, mediating them through restricted gateways, VPN-only access, or network-level allowlists; adopt install-time exposure testing with automated warnings when management interfaces become publicly reachable~\cite{NREL_standards}. Because most exposed deployments are generic or installer-layered rather than manufacturer-branded (Section~\ref{sec:hetero}, Table~\ref{tab:rq1_brand_matches}), this is where much of the attack surface can be removed.
\item \textbf{Operators:} monitor for anomalous DER command patterns, such as coordinated disconnects or bulk active/reactive setpoint changes, characteristic of the manipulation attacks modeled in Section~\ref{sec:power-analysis}. Because a large fraction of devices persist across snapshots (Section~\ref{sec:persistence}), such monitoring must be sustained over time.
\item \textbf{Regulators/standards bodies:} much of the observed exposure is introduced by installers and third-party integrators, not by manufacturer defaults (Section~\ref{sec:hetero}); standards that reach these parties, mandating secure-by-default deployment and disclosure of Internet-reachable management interfaces, would close a gap that manufacturer-focused requirements alone cannot.
\end{itemize}}
Reducing reachability of these surfaces also mitigates CVE risk in management services (Section~\ref{sec:vulns}, Table~\ref{tab:rq2_cve_summary}) even before patches propagate. \new{These defenses act by reducing the compromisable device set, whose grid-impact effect we quantify in Section~\ref{sec:res}. Operators also deploy protective schemes such as undervoltage load shedding, but these are designed for natural disturbances, not adversarial manipulation: an attacker aware of relay settings could potentially craft trajectories that avoid triggering protection while still causing damage. Modeling such protection mechanisms against adversarial manipulation is an important direction for future work.}

\paragraph{\textbf{Grid Impact at Today's Exposure Levels.}} Our measurement-grounded grid analysis (Section~\ref{sec:power-analysis}) shows that the exposed inverter population is operationally consequential at present-day deployment levels. Coordinated manipulation of 41.1~MW can drive nearly half of the Oahu network's buses below safe voltage (Figure~\ref{fig:voltage-results}) and overloads 6.7\% of transmission lines up to 12\% beyond nominal ratings (Figure~\ref{fig:flow-results}). Such conditions induce significant thermal stress on conductors, transformers, and motors and could force protective disconnections that propagate into widescale outages.

Two features make this risk fundamentally different from prior cyber-physical attacks. First, (power) supply-side compromise gives attackers direct control over both active and reactive power, enabling voltage manipulation that demand-side attacks~\cite{shekari2022madiot,soltan2018blackiot,rodriguez2025madiot} cannot achieve, explaining why orders of magnitude fewer devices are sufficient. \new{Demand-side attacks such as MaDIoT, by contrast, require coordinating millions of devices to achieve comparable impact.} Second, our analysis is grounded in inverters exposed today rather than hypothetical botnet populations, shifting the question from \emph{``what could happen if attackers controlled X devices?''} to \emph{``what can attackers do with the X devices already exposed?''} As DER penetration grows, this gap between Internet exposure and grid risk will widen unless deployment practices evolve~\cite{dersec_2025,jablanski_energy,NREL_standards}.

\paragraph{\textbf{\new{Limitations.}}} Our analysis has several limitations. Geographic attribution relies on IP-based geolocation from Censys, which may not reflect the physical location of deployed devices due to VPNs, cloud relays, or backhauled monitoring infrastructure. Our measurement captures only DER devices directly reachable from the public Internet, representing the exposed attack surface rather than the full installed base. Device characterization depends on scanner-visible artifacts; devices that obfuscate this information may be undercounted. CVE associations are derived from disclosed software versions and do not imply confirmed exploitability.

The two-snapshot design introduces additional considerations. The LLM-based labels used for training (Stage~2) may contain errors; we mitigate this through deterministic refinement rules and manual validation, achieving 98\% label accuracy. The Random Forest must also generalize from Snapshot~1 to Snapshot~2 across a four-month gap and a nearly threefold expansion in vendor coverage; we address this through features capturing general DER characteristics rather than brand-specific patterns, and through consensus voting and threshold tuning. Finally, short brand-name matching uses word-boundary constraints to reduce spurious substring matches, though some coincidental overlaps may remain.

\paragraph{\textbf{\new{Future Work.}}} Future work includes extending the classifier to track exposure longitudinally across additional snapshots and improving attribution for generic deployments. Complementary user studies of installers and service providers could clarify why insecure configurations persist in practice~\cite{raymaker2025sea}. On the power system side, \new{our sensitivity analysis (Section~\ref{sec:res}) quantifies impact under a reduced compromised set; extending this to incorporate further adversarial constraints, such as imperfect topology knowledge or coordination timing, would refine how measured exposure translates into grid risk. Modeling protection mechanisms (e.g., undervoltage load shedding) and their interaction with adversarial manipulation is another important direction.} Finally, \new{dynamic power-system analysis of attacks over shorter time scales, such as frequency instability, remains open and} contrasts with our steady-state analysis.

\section{Acknowledgments}
We thank the anonymous reviewers for their constructive feedback. This material is based upon work supported by the U.S. Department of Energy's Office of Cybersecurity, Energy Security, and Emergency Response (CESER) under Award DE-CR0000056, the National Science Foundation's Secure and Trustworthy Cyberspace (SaTC) program under Award 2231651, and the NSF Graduate Research Fellowship under Grant DGE-2039655.

\bibliographystyle{ACM-Reference-Format}
\bibliography{refs}

\appendix %

\section{Open Science}
This work involves Internet-scale measurement of security-relevant exposure in deployed critical infrastructure. We do not publicly release the discovery pipeline, Internet-wide queries, or host-level datasets, as these artifacts could be misused to locate and target vulnerable solar DER deployments. The paper provides a detailed description of the measurement and analysis methodology to support independent evaluation, and selected artifacts may be shared with vetted researchers upon request. The power system analysis code, including the adversarial OPF formulation and the augmented Oahu test case, is publicly available at~\cite{power_repo}. A variation of the analysis code is also publicly displayed at \href{https://gridtrouble.xyz}{https://gridtrouble.xyz}.

\section{Ethical Considerations}
\label{app:ethics}
\paragraph{\textbf{Stakeholders.}}
Stakeholders affected by this work include solar DER owners and operators, installers and service providers, inverter and platform vendors, grid operators and utilities, regulators, and the broader public that depends on reliable and secure power systems. Additional stakeholders include vendors whose products and default configurations are implicated, as well as researchers and practitioners studying critical infrastructure security.

\paragraph{\textbf{Impacts and Potential Harms.}}
This study analyzes Internet-exposed solar DER platforms using publicly available Internet scan data and scanner-visible artifacts such as service banners, web interfaces, and protocol metadata. We did not interact with devices beyond passive observation, did not authenticate to systems, and did not modify device state. Potential harms include the risk that characterizing exposed interfaces and vulnerable software versions could be misused or could cause concern among vendors and operators. In addition, the presence of a vulnerable software version does not necessarily imply exploitability in a given deployment.

\paragraph{\textbf{Mitigations.}}
To mitigate potential harms, we analyze exposure at an aggregate level and do not disclose IP addresses, credentials, or exploit details. Vulnerability findings are framed as potential exposure to known CVEs based on scanner-visible information, and we explicitly avoid claims of active exploitation. We also pursue responsible disclosure by notifying affected vendors of exposure trends and common misconfigurations prior to publication, with the goal of supporting remediation and more secure deployment practices. Several vendors have acknowledged these disclosures, and we are engaged in ongoing technical discussions aimed at improving deployment practices. Beyond vendor disclosure, we have briefed senior leadership at the U.S. Department of Energy's Office of Cybersecurity, Energy Security, and Emergency Response (CESER), which oversees grid cybersecurity, on the outcomes of this research to support broader grid-security efforts.

Our companion tool (\href{https://gridtrouble.xyz}{https://gridtrouble.xyz}) simulates and visualizes potential grid-level attacks on Oahu using the actual exposed inverter population from our measurement. During review, the tool is password-protected with credentials shared only with reviewers (password: mz692Am1bsfdrXBO). At publication, the publicly accessible version will use a sanitized grid model that preserves qualitative attack patterns for educational and research use but cannot be used to plan an attack on the actual Oahu network; access to the unmodified model will be restricted to vetted researchers.

\paragraph{\textbf{Ethical Decision.}}
We determined that proceeding with this research and its publication is ethically justified because it relies solely on passive measurement of publicly reachable systems, does not violate reasonable expectations of privacy or consent, and provides societal benefit by improving visibility into an emerging attack surface affecting critical energy infrastructure.

\section{Query Construction Details}
\label{app:query_details}

\noindent\textbf{Scanner-Visible Field Matching.}
Each query is generated from a fixed template that searches scanner-visible fields where deployment identifiers are likely to appear.  In addition to HTTP response titles and HTML bodies, we match fingerprints against Censys-extracted metadata including TLS certificate subject fields and service-level response data.  This allows queries to
capture identifiers exposed through device web interfaces, monitoring portals, embedded management servers, and certificate configurations.

These fields are selected because deployed DER systems frequently surface product names, model identifiers, portal domains, or vendor branding in at least one scanner-visible artifact, even when branding is absent from the primary web interface. Manual inspection of exposed devices confirms that identifiers may appear inconsistently across titles, page content, certificate subjects, or auxiliary service metadata, making multi-field matching necessary for robust discovery.

For vendors with distinctive identifiers, such as \emph{Sunny Boy}, \emph{IQ Gateway}, or \emph{SUN2000}, fingerprints are incorporated directly as exact or near-exact string matches.  For vendors with common or ambiguous names (e.g., ABB, Delta, Siemens), we apply additional constraints requiring solar- or photovoltaic-specific context to reduce false positives from unrelated industrial or corporate infrastructure. In practice, this is implemented by conjoining vendor identifiers with
energy-domain terms such as \texttt{solar}, \texttt{inverter}, or \texttt{photovoltaic}, ensuring that matches reflect DER-related deployments and not unrelated web services.

\noindent\textbf{Generic Discovery Query.}
The generic discovery query is derived from the same vendor
documentation corpus used for fingerprint extraction, by identifying DER-specific interface terms that consistently appear across vendors once brand names are removed.  Concretely, we extract recurring phrases associated with solar inverter monitoring and control (e.g., \texttt{photovoltaic system}, \texttt{solar inverter}, \texttt{energy monitoring}, \texttt{power output}, \texttt{solar dashboard}) and combine them into a single query without vendor identifiers.  While less precise than brand-specific queries, this approach targets interface language that is characteristic of deployed solar DER systems and expands coverage to devices that do not expose recognizable branding but still present DER-relevant interfaces on the public Internet.

\noindent\textbf{Document Mining.}
For each vendor, we scrape publicly accessible installation manuals, datasheets, and support documentation.  These sources include inverter hardware vendors (e.g., SMA, Fronius, Huawei), monitoring and logging platforms (e.g., Solar-Log, Solarman, eGauge), and hybrid systems that combine control and telemetry. Documents are parsed from PDF and HTML formats into plain text. We extract identifiers that vendors explicitly reference as part of device configuration, monitoring, or management workflows. For example, SMA documentation repeatedly names product families such as \emph{Sunny Boy} and \emph{Sunny Tripower}, model identifiers such as \texttt{SB3000}, and web-facing components such as \emph{Sunny WebBox} and account login endpoints.

Identifiers are extracted deterministically using rule-based parsing: (i)~product names and model numbers specified in configuration or status interfaces, (ii)~user-interface strings that appear in documented dashboards or login pages, and (iii)~portal domains and service names used for remote monitoring or fleet management.  Extracted identifiers are consolidated through exact matching and edit-distance clustering to merge minor textual variants (e.g., capitalization or spacing differences).  The resulting fingerprints are grouped by vendor and used directly in downstream query construction.

\section{LLM Classification and Refinement Details}
\label{app:llm_refinement}

\begin{figure*}[]
    \centering
    \includegraphics[width=\textwidth]{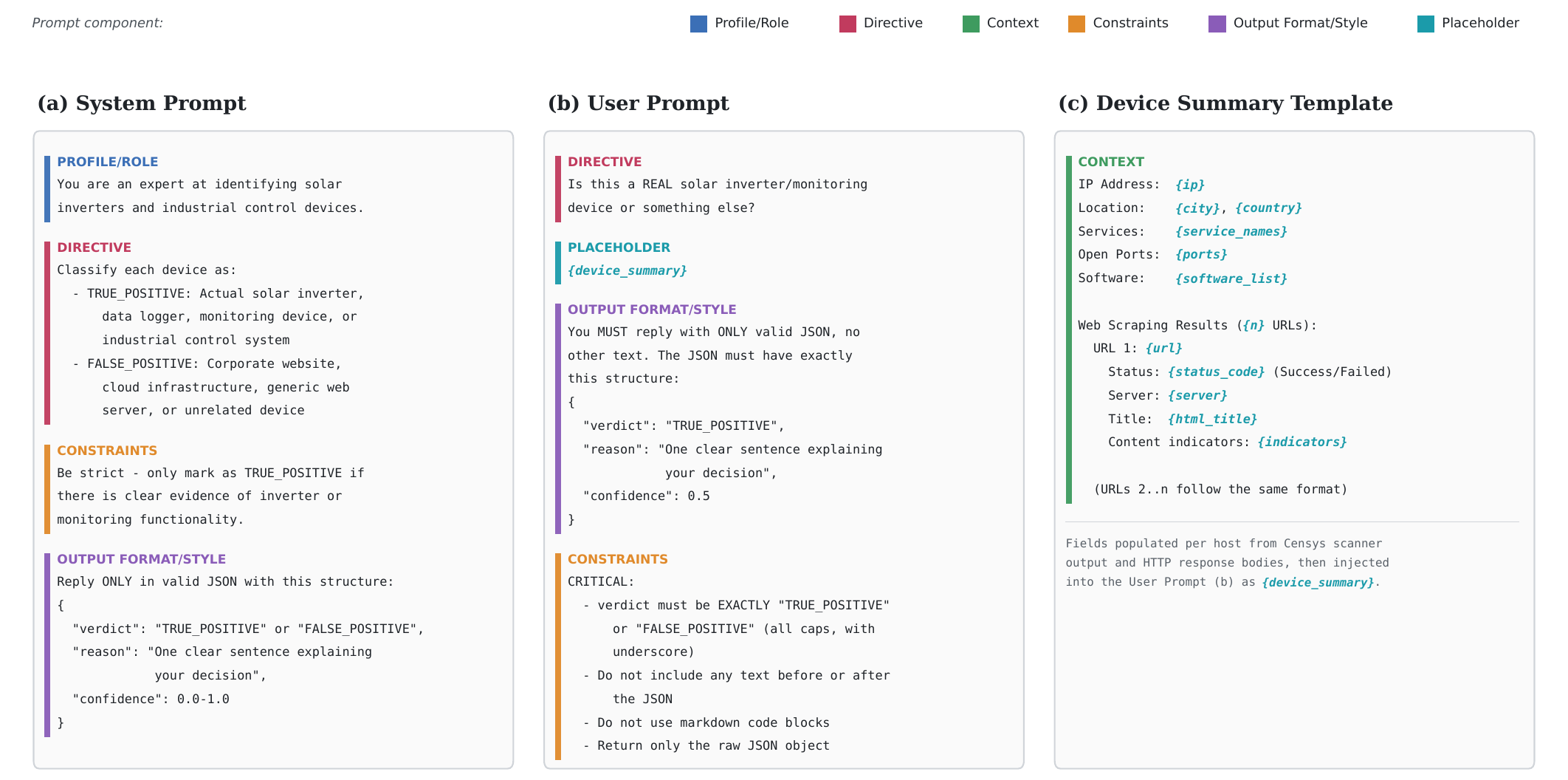}
    \caption{\textbf{Three-component structure of the LLM classification prompt.} (a)~The system prompt defines the classifier role, the binary verdict space, and a strict JSON output schema.  (b)~The user prompt instantiates a single classification query and reiterates formatting constraints, inserting the per-host \texttt{\{device\_summary\}} at runtime. (c)~The device summary template is populated with scanner-visible evidence from Censys, including service metadata, open ports, software, and up to $n$
    scraped web responses.  Placeholder fields (rendered in teal italics) are substituted per host before the template is injected into panel~(b).}
    \label{fig:prompt_structure}
\end{figure*}

\noindent\textbf{Prompt Structure.}
Host classification is performed by a structured three-component prompt illustrated in Figure~\ref{fig:prompt_structure} that separates invariant logic from per-host evidence. The system prompt~(a) fixes the classifier role and a strict JSON schema.  The user prompt~(b) issues the query and injects a per-host \texttt{\{device\_summary\}}, which the template in~(c) populates from Censys scanner output.  Only the device summary varies across the $\sim$166K candidate hosts, keeping the screening logic fixed across all models in Table~\ref{tab:llm_comparison}.

\noindent\textbf{LLM Evaluation Protocol.}
To characterize LLM behavior and select an appropriate screening model, we evaluated multiple LLMs with different architectures and context lengths. Table~\ref{tab:llm_comparison} summarizes performance based on
manual inspection of randomly sampled outputs.  For each model, we uniformly sampled 50~hosts classified as DERs and 50~classified as non-DERs, and manually assessed correctness to estimate relative precision and recall for screening purposes.

Across all evaluated models, misclassifications are dominated by false positives rather than false negatives.  For example, Llama3.3:70b and Qwen3:32b (8K) each exhibit only 1~false negative in 50~non-DER samples while maintaining a high true negative rate.  This behavior reflects the intended role of the LLM as a conservative screening filter that substantially reduces the non-DER search space while preserving the majority of true DER deployments.

\noindent\textbf{Model Selection.}
While the evaluated models exhibit broadly similar error characteristics, we select Llama~3.3:70b as the base classifier. Llama identifies a larger number of candidate DER hosts than the Qwen-based models while exhibiting comparable true positive and negative rates. Retaining a larger candidate set at this stage is desirable because subsequent steps are designed to conservatively eliminate residual false positives.  In practice, applying Llama~3.3:70b reduces the candidate set from 166K to 67K hosts.

\noindent\textbf{Systematic Error Analysis.}
Manual inspection of misclassified samples across all evaluated models revealed that classification errors were not random but clustered around a small number of recurring patterns.  Common false positives included:

\begin{itemize}[nosep]
\item \textbf{Consumer-grade solar-powered cameras} (e.g., Reolink): these hosts use terms such as ``solar,'' ``battery,'' and ``power'' despite having no grid connection or inverter functionality. They exhibit a consistent signature including Reolink-branded web interfaces, camera-specific services, and the absence of inverter sensor information or control endpoints.

\item \textbf{Network management appliances}: devices with web dashboards exposing power or energy metrics in a network monitoring context rather than a DER context.

\item \textbf{Academic and institutional web pages}: pages discussing power systems that also expose industrial protocols (e.g., Modbus) without operational control surfaces.
\end{itemize}

False negatives, while rare, typically corresponded to genuine DER devices that lacked explicit branding or used non-standard naming conventions.

\noindent\textbf{Suppression Rules.}
From false-positive samples, we extract recurring scanner-visible characteristics, including HTML titles, server headers, software identifiers, protocol combinations, and vendor or product strings, and encode these as explicit suppression rules that remove known non-DER artifacts. For example, the Reolink suppression rule removes hosts whose scanner-visible artifacts match Reolink-specific identifiers without co-occurring DER control or monitoring features. Similar rules are constructed for other recurring false-positive classes.

\noindent\textbf{Final Validation.} After refinement, we perform a final round of manual validation on a randomly sampled subset of the resulting DER host set. This confirms a
true positive rate of 98\% and a true negative rate of 100\% for the final refined dataset, as summarized in Table~\ref{tab:classification_validation}.  The remaining misclassification corresponds to a single academic website discussing power systems that exposed an industrial control protocol but did not represent an operational DER deployment.  Given its ambiguity and lack of control functionality, we conservatively exclude this host from subsequent analysis.
\begin{table}[t]
  \centering
  \small
  \caption{Stage~2 classification pipeline. LLM screening reduces non-DER artifacts with high recall; filtering suppresses recurring false-positive patterns, producing the final labeled dataset used to train the Stage~3 classifier.}
  \setlength{\tabcolsep}{6pt}
  \renewcommand{\arraystretch}{1.15}
  \begin{tabular}{lccc}
    \toprule
    \textbf{Stage} &
    \textbf{Hosts retained} &
    \textbf{$^\dagger$P} &
    \textbf{$^\circ$R} \\
    \midrule
    LLM screening &
    67,293 &
    58\% &
    97\% \\
    Filtering &
    41,755 &
    98\% &
    100\% \\
    \bottomrule
  \end{tabular} \\
   \footnotesize
  $^\dagger$P = precision; $^\circ$R = recall\\
  \label{tab:classification_validation}
  \vspace{-1em}
\end{table}

\section{Probability Threshold Selection}
\label{app:threshold}

\begin{figure}[]
    \centering
    \includegraphics[width=\linewidth]{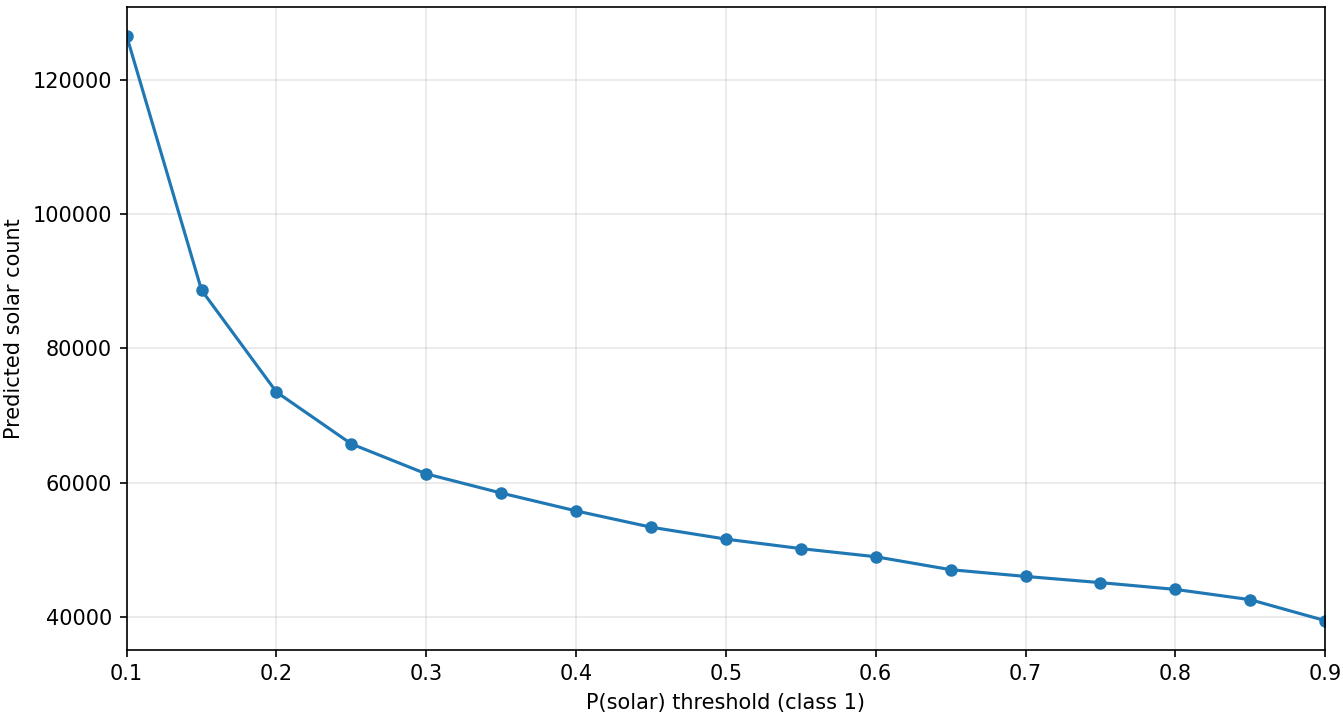}
    \caption{\textbf{Predicted DER host count as a function of the solar probability threshold.}  The count drops steeply at low thresholds and stabilizes above 0.25.  We select the elbow at 0.25 as the operating point.}
    \label{fig:threshold_sweep}
\end{figure}

The Random Forest classifier produces a probability estimate
$P(\text{solar})$ for each host.  To select a classification threshold, we sweep values from 0.10 to 0.90 in increments of 0.05. Figure~\ref{fig:threshold_sweep} shows the predicted DER host count at each threshold.  We select 0.25, the elbow of the curve, where the predicted count transitions from steep decline to gradual decrease. Manual inspection of a random sample of  100~hosts at this threshold yields a true positive rate of 99\%.

This threshold is applied in conjunction with five-run consensus voting (Section~\ref{sec:scaled_classification}), which further removes hosts whose classification is unstable across training perturbations.

\section{Geographic Distribution of Exposed DERs}

\label{app:geography}

\begin{figure}[]
    \centering
    \includegraphics[width=0.95\linewidth]{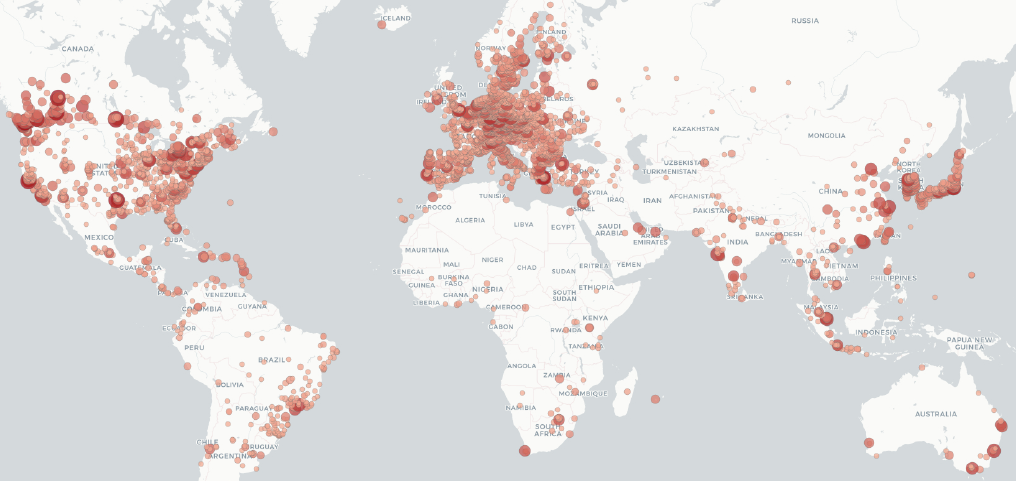}
    \caption{Global distribution of Internet-exposed solar DER hosts. Each point represents a geolocated Internet-facing interface associated with a validated solar DER platform.}
    \label{fig:global_der_map}
\end{figure}

We show the geographic distribution of Internet-exposed solar DER hosts in Figure~\ref{fig:global_der_map}. We can see that DER-exposure is global but highly concentrated. Canada alone accounts for 24.2\% of all exposed hosts (16,050), followed by the United States at 18.6\% (12,332); together, these two countries comprise nearly half of all observed exposure. A second tier of countries contributes a substantial additional share, including Germany (6.9\%), France (4.9\%), and Japan (3.9\%), while the remaining top ten countries each contribute between 2.5--3.4\%. Countries central to recent grid-security incidents also appear in our dataset, including Poland (1.1\%, 729 hosts) and Ukraine (0.3\%, 176 hosts), though at smaller scale. Table~\ref{tab:rq1_country_exposure} (in the appendix) reports the geographic distribution for the top countries in our dataset.

Beyond these regions, exposure extends broadly across the globe. A long tail of 147 additional countries collectively contributes 13.5\% of exposed hosts (8,937), indicating that Internet-exposed solar DER platforms are not confined to a small number of grids, markets, or regulatory regimes.

In contrast to this broad global presence, exposure in China appears comparatively limited given its dominant role in the global solar industry. Despite being the world’s largest solar PV manufacturer and installer, China accounts for only 2.7\% of exposed hosts (1,777) in our dataset~\cite{iea_2025}. This discrepancy likely reflects differences in deployment and connectivity practices. Many Chinese inverters and monitoring systems are deployed behind carrier-grade NATs, operate on private or utility-managed networks, or rely on cloud-mediated control channels that do not expose device-facing interfaces directly to the public Internet. In addition, regulatory and operational norms in China may discourage direct Internet exposure of device management interfaces.

Interpreting these counts requires care. Host geolocation reflects the network location of the Internet-facing interface, which may differ from the physical location of the underlying DER, for example due to cloud-hosted monitoring portals, ISP backhaul, or centralized installer infrastructure. Even with this caveat, \textit{two security-relevant patterns} emerge. First, exposure broadly aligns with regions that have high penetration of residential and small-commercial PV and rely heavily on installer-managed monitoring platforms, suggesting that Internet exposure commonly arises from standard deployment and management practices~\cite{irena_2024}. Second, the substantial long tail indicates that insecure exposure appears across diverse markets and regulatory environments, pointing to a systemic issue. Taken together, the geographic distribution shows that Internet exposure of solar DER platforms is a global phenomenon shaped by how DERs are deployed and managed, implying that effective mitigation will require changes to installation and operational practices at scale.

\section{Extra Tables}
\begin{table}[t]
\centering
\small
\caption{Geographic distribution of internet-exposed solar DER hosts (top 20 countries).}
\begin{tabular}{lrr}
\toprule
\textbf{Country} & \textbf{Exposed hosts} & \textbf{\% total} \\
\midrule
Canada               & 16,050 &  24.2\% \\
United States        & 12,332 &  18.6\% \\
Germany              &  4,563 &   6.9\% \\
France               &  3,277 &   4.9\% \\
Japan                &  2,576 &   3.9\% \\
Brazil               &  2,279 &   3.4\% \\
Italy                &  2,086 &   3.1\% \\
Netherlands          &  1,949 &   2.9\% \\
China                &  1,777 &   2.7\% \\
South Korea          &  1,632 &   2.5\% \\
Greece               &  1,561 &   2.4\% \\
Sweden               &  1,414 &   2.1\% \\
Spain                &  1,190 &   1.8\% \\
Czech Republic       &    963 &   1.5\% \\
Austria              &    738 &   1.1\% \\
Poland               &    729 &   1.1\% \\
United Kingdom       &    628 &   0.9\% \\
India                &    598 &   0.9\% \\
Turkey               &    564 &   0.8\% \\
Portugal             &    535 &   0.8\% \\
All other countries (147) & 8,937 & 13.5\% \\
\bottomrule
\end{tabular}
\label{tab:rq1_country_exposure}
\end{table}

\FloatBarrier

\begin{table*}[t]
\centering
\small
\caption{Solar DER vendors by device and platform role. Vendors without a dagger (\dag) are included in both Snapshot~1 (34~vendors) and Snapshot~2. Vendors marked \dag\ are added in Snapshot~2 (62~additional vendors, 96~total).}
\setlength{\tabcolsep}{4pt}
\begin{tabular}{@{}llll@{}}
\toprule
\textbf{Vendor} & \textbf{Type(s)} & \textbf{Vendor} & \textbf{Type(s)} \\
\midrule
ABB                     & Inverter                   & Meteocontrol\dag         & Monitoring platform \\
Afore\dag               & Inverter                   & Midnite Solar\dag        & Charge controller \\
Also Energy\dag         & Monitoring platform        & Morningstar\dag          & Charge controller \\
Altenergy\dag           & Microinverter              & NEP\dag                  & Microinverter \\
APsystems\dag           & Microinverter              & Neurio\dag               & Energy meter \\
AUO\dag                 & Microinverter              & Omnik\dag                & Inverter \\
Beckhoff\dag            & Industrial automation      & Omnidian\dag             & Monitoring platform \\
Bosswerk\dag            & Inverter                   & Outback Power            & Inverter \\
Brookfield Renewable    & Fleet management           & Power Electronics        & Utility-scale inverter \\
Campbell Scientific     & Data logger                & PowerHub\dag             & Inverter; Monitoring \\
Canadian Solar\dag      & Inverter                   & Q~Cells\dag              & Inverter \\
Carlo Gavazzi\dag       & Energy meter               & Refu                     & Inverter \\
CHINT\dag               & Inverter                   & Revolt\dag               & Inverter \\
Contec                  & Monitoring gateway         & SAJ\dag                  & Inverter \\
Curb Energy\dag         & Energy monitor             & Samil\dag                & Inverter \\
Danfoss\dag             & Inverter                   & Satel Iberia\dag         & Monitoring gateway \\
Delta                   & Inverter                   & Schneider Electric       & Inverter; Energy mgmt \\
Deye                    & Inverter                   & Sense\dag                & Energy monitor \\
Digi International\dag  & IoT gateway                & Shanghai Hi-Flying\dag   & WiFi module (inverter) \\
eGauge                  & Energy meter               & Shoals Technologies\dag  & Combiner/monitoring \\
Emporia Energy\dag      & Energy monitor             & Siemens                  & Inverter; Energy mgmt \\
Enphase                 & Microinverter; Portal      & Sineng Electric          & Inverter \\
EPEver\dag              & Charge controller          & Skyfri\dag               & Monitoring platform \\
Fimer (ABB Solar)       & Inverter                   & SMA                      & Inverter; Portal \\
Fox ESS\dag             & Inverter                   & Smappee\dag              & Energy monitor \\
Fronius                 & Inverter; Portal           & Sofar Solar\dag          & Inverter \\
FusionSolar             & Monitoring portal          & Sol-Ark\dag              & Inverter \\
GE Vernova\dag          & Utility-scale inverter     & Solar Analytics\dag      & Monitoring platform \\
Generac\dag             & Inverter; Battery          & Solar-Log                & Monitoring gateway \\
GivEnergy\dag           & Inverter; Battery          & SolarBOS\dag             & Combiner/monitoring \\
GoodWe                  & Inverter; Portal           & SolarEdge                & Inverter; Portal \\
Growatt                 & Inverter; Portal           & Solarman                 & Monitoring platform \\
Hanwha\dag              & Inverter                   & Solax                    & Inverter; Portal \\
Hitachi Energy\dag      & Utility-scale inverter     & Solectria\dag            & Inverter \\
Hoymiles                & Microinverter; Portal      & Solivia                  & Inverter \\
Huawei                  & Inverter; Portal           & Solis (Ginlong)          & Inverter; Portal \\
IBC Solar\dag           & Inverter                   & Span\dag                 & Smart panel \\
Inavitas\dag            & Monitoring platform        & Sungrow                  & Inverter; Portal \\
Ingeteam\dag            & Inverter                   & SunPower\dag             & Inverter; Portal \\
Kostal\dag              & Inverter                   & Sunsynk\dag              & Inverter \\
Kstar\dag               & Inverter                   & TBEA                     & Inverter \\
Locus Energy            & Monitoring platform        & Tesla\dag                & Inverter; Battery \\
LoggerNet               & Data logging software      & Tigo                     & Optimizer; Monitoring \\
SEL\dag                 & Protection relay           & Trannergy\dag            & Inverter \\
                        &                            & Victron                  & Inverter; Energy mgmt \\
                        &                            & Weidmuller\dag           & Monitoring gateway \\
                        &                            & Yaskawa\dag              & Inverter \\
                        &                            & ZCS Azzurro\dag          & Inverter \\
                        &                            & Zeversolar\dag           & Inverter \\
\bottomrule
\end{tabular}
\label{tab:brand_types}
\end{table*}

\FloatBarrier

\end{document}